\documentclass[12pt]{iopart}

\usepackage{iopams}
\usepackage{graphicx}
\usepackage{float}
\usepackage{color}

\newcommand{\rem}[1]{}

\newcommand{\revision}[2]{#2}

\begin{document}

\newtheorem{theorem}{Theorem}[section]
\newtheorem{definition}[theorem]{Definition}
\newtheorem{lemma}[theorem]{Lemma}
\newtheorem{proposition}[theorem]{Proposition}
\newtheorem{rema}[theorem]{Remark}

\def\boldeta{\boldsymbol{\eta}}

\newcommand{\boldr}{\boldsymbol{r}}
\newcommand{\mse}{\mathfrak{se}}
\newcommand{\de}{\delta}
\newcommand{\boldphi}{\boldsymbol{\phi}}
\newcommand{\boldpsi}{\boldsymbol{\Psi}}
\newcommand{\boldalpha}{\boldsymbol{\alpha}}
\newcommand{\boldbeta}{\boldsymbol{\beta}}
\newcommand{\bb}{\boldsymbol{b}}
\newcommand{\mg}{\mathfrak{g}}
\newcommand{\bv}{\boldsymbol{v}}
\newcommand{\bu}{\boldsymbol{u}}

\title[Helical states of nonlocally interacting molecules]{Helical states of nonlocally interacting molecules and their linear stability: geometric approach}

\author{S Benoit$^{1}$, D D Holm$^{2,\,3}$ and V Putkaradze$^{1,\,4}$}
\address{$^1$ Department of Mathematics, Colorado State University, Fort Collins, CO 80523 USA}
\address{$^2$ Department of Mathematics, Imperial College London, London SW7 2AZ, UK}
\address{$^3$ Institute of of Mathematical Sciences, Imperial College London, London SW7 2AZ, UK}
\address{$^4$ Department of Mechanical Engineering, University of New Mexico, Albuquerque NM 87131 USA}
\ead{benoit@math.colostate.edu}

\rem{
\author{Steve Benoit$^{1}$, Darryl D. Holm$^{2,\,3}$ and Vakhtang Putkaradze$^{1,\,4}$
\vspace{2mm}
\\
{\small \!\!\!$^1$ \it Department of Mathematics, Colorado State University, Fort Collins, CO 80523 USA}\\
{\small \!\!\!$^2$ \it Department of Mathematics, Imperial College London, London SW7 2AZ, UK}\\
{\small \!\!\!$^3$ \it Institute of of Mathematical Sciences, Imperial College London, London SW7 2AZ, UK}\\
{\small \!\!\!$^4$ \it Department of Mechanical Engineering, University of New Mexico, Albuquerque NM 87131 USA}\\
\\ \\}
}

\begin{abstract}
The equations for  \revision{R}{strands} of rigid charge configurations interacting nonlocally are formulated on the special Euclidean group,  $SE(3)$, which naturally generates helical conformations. Helical stationary shapes are found by minimizing the energy for rigid charge configurations positioned along an infinitely long molecule with charges that are off-axis. The classical energy landscape for such a molecule is complex \revision{R}{with a large number of energy minima}, even when limited to helical shapes.  The question of linear stability and selection of stationary shapes is studied \revision{R}{using} an $SE(3)$ method that naturally accounts for the helical geometry.  We investigate the linear stability of a general helical polymer that possesses torque-inducing non-local self-interactions and find the exact dispersion relation for the stability of the helical shapes with an arbitrary interaction potential. We explicitly determine the linearization operators and compute the numerical stability for the particular example of a linear polymer comprising a \revision{R}{flexible rod with a repeated configuration of two equal and opposite off-axis} charges, thereby showing that even in this simple case the non-local terms can induce instability \revision{G}{that leads to the \revision{R}{rod} assuming helical shapes}.
\end{abstract}


\submitto{\JPA}

\maketitle 
 
\section{Introduction}
Molecules with repeating subunits that spontaneously form helical shapes are ubiquitous in nature. A classical example is given by $\alpha$-helices, which are an important motif in the secondary structure of proteins \cite{BrTo1999}. Other examples of naturally occurring helical structures include intermediate filament proteins keratin and vimentin \cite{BuStSt2001}, the myosin and kinesin families of motor proteins \cite{BuStSt2001}, tubulin microtubules \cite{LiDeNiNoDo2002}, RecA and Rad51 filaments on DNA \cite{YuJaWeOgEg2002}, flagella and pili \cite{KJ2009}, and others.  Artificial polymer structures that spontaneously take helical forms have also been obtained, see for example \cite{Co-etal-2001,MaGoScMe2005}.
The spontaneous emergence and persistence of helical shapes has been the subject of several recent studies. For progress in the description of helical shapes based on elastic rod theory,  see \cite{ChGoMa2006}. The spontaneous formation of helical structures under compaction of an elastic rod has also been shown, by using a Lennard-Jones-like potential whose repulsive core eliminates self-intersections \cite{Ba-etal-2007}.

Molecular shapes arise through an interplay between elastic forces caused by bending and twisting of the molecular bonds, and long- and short-range forces (such as electrostatic and van der Waals forces) between individual atoms comprising the molecule.  For elastic interactions only, Kirchoff's rod theory \cite{Di1992,DiLiMa1996} has been used to model molecular shapes with notable success \cite{GoTa1996, GoPoWe1998,BaMaSc1999,GoGoHuWo2000, HaGo2006, ChGoMa2006,NeGoHa2008}.  The recent work on non-local interactions has considered helical molecules as rods with charges concentrated along the axis of the rod only \cite{DiLiMa1996, ChMa2004, ChGoMa2006, DeLi2009}. Other authors used energy optimization \revision{R}{with} short-range repulsion only, without initial \revision{R}{assumptions} on the molecular shapes, and yet helical shapes showed surprising persistence \cite{Ba-etal-2007,Ma-etal-2000}, either for the whole molecule, or in parts of it. The crucial question is whether such helical shapes persist for more complex interactions or geometries of the molecules.
For example, more accurate models of molecules could include charges that are attached at a certain distance from the axis, as shown in Fig. \ref{fig:sketch}. Interatomic forces are caused by some potential -- or combination of potentials -- that depends explicitly on the \revision{R}{geometry of the molecule and the} Euclidean distance $d$ between two points $s$ and $s'$ along the curve representing the axis of the molecule. Equations of motion for such molecules were derived recently using nonlocal extensions of exact geometric rod theory \cite{HoPu2008,El-etal-2009}.  However, the question of existence of helical shapes for more complex interactions between the parts of the molecule and the methods for their computations has remained open. In this paper, we demonstrate a fast and efficient method for finding such states, and in addition show  how we can achieve a complete classification of all helical states. We also show that the geometric approach allows exact computation of dispersion relations and linear instability growth rates. 

\paragraph{Plan of the paper.} Section \ref{derivation-sec} derives the equations for nonlocally interacting \revision{R}{rods} of rigid charge configurations, \revision{Q1}{called \emph{bouquets}. These bouquets are fixed configurations of  charges, rigidly attached to the molecule's central axis.  The molecule is comprised of a long repeating chain of such bouquets.} The equations are formulated on the special Euclidean group, $SE(3)$, consisting of three-dimensional spatial rotations and translations, which naturally generate helical configurations. Section \ref{linearstrand-sec} demonstrates how to find helical stationary shapes by minimizing the energy for bouquets positioned along  an infinitely long molecule with charges that are off-axis.   As shown in Section \ref{sec:minenergy}, even this simple molecule, when deformed into helical configurations, shows quite a complex and intriguing energy landscape. Section \ref{multi-helices-sec} briefly discusses how to extend the ideas presented above to include more general shapes. In particular, we consider a molecular shape \revision{Q1}{we refer to as a 2-helix.  Here, we define an $n$-helix as a molecule that is a true helix consisting of repeating groups of $n$ bouquets along the axis.} For a 2-helix, one optimizes the energy over both the shape of the helix and the relative configuration of two \revision{R}{bouquets}. One may treat $n$-helices by optimizing configurations of $n$ bouquets using the same methods.  After revealing the complexity of the energy landscape, even when limited to helical shapes, and the large number of energy minima in that ``helical universe'', it is natural to pose the question of linear stability and selection of the stationary shapes. By using an $SE(3)$ method that naturally accounts for the helical geometry, in Section \ref{linstability-sec} we investigate the linear stability of a general helical polymer that possesses torque-inducing non-local self-interactions. As far as we know, no such work has been undertaken previously. The geometric approach in Section \ref{linstability-sec} allows us to find the exact dispersion relation for the stability of the helical shapes with an arbitrary interaction potential. Of course, for particular applications it is important to explicitly compute the linearization operators. This is accomplished in Section \ref{comp-derivs-sec} \revision{R}{using geometric methods to compute} derivatives of the potential energy. Section \ref{num-stability-linear-sec} computes the numerical stability for the particular example of a linear polymer comprising a charged \revision{R}{rod} with repeated configuration of two equal and opposite charges that interact through a screened electrostatic and Lennard-Jones potential. For the sake of simplicity, we concentrate on the stability of a polymer that is perfectly straight in its unstressed configuration. Such a polymer is neutrally stable in the absence of the nonlocal interactions. It is therefore interesting that nonlocal terms can induce instability \revision{G}{that causes the molecule to deform into helical conformations}. Physically, this instability is connected to the tendency of the \revision{R}{rod} to minimize its energy and properly align the dipole moments of each bouquet by twisting, as was already seen in the minimum energy calculation shown in Section \ref{sec:minenergy}.

\revision{Q6}{We shall note that these results are difficult to achieve in the traditional Kirchhoff-based approach, as the non-local interactions depend on the relative distance and orientation of the charges at different points on the rod. In these traditional methods, the equations of motion are written in a coordinate system that is moving with the rod, and changing with both position on the rod $s$ and time $t$.  Thus, one has to write equations of motion, \emph{i.e.} calculate the momenta and forces, in a frame attached to the rod that is moving and rotating in a non-inertial fashion.  The main difficulty arises because the relative distance between the charges depends on both their bouquet positions on the centerline and the local rotation of the bouquet about the centerline.  To find this orientation one needs to integrate auxiliary equations of motion at each step (\emph{Darboux}'s vector), so the solution must be known \emph{before} the relative distance between the charges can be computed. However, the solution explicitly depends on the relative distance between the charges, and so the closure of the system is problematic.  These complications make an explicit derivation of equations of motion from Kirchhoff's approach difficult, if not impossible.  On the other hand, the approach we suggest here provides a very straightforward derivation.  One need not deal with vectors, forces and torques in non-inertial frames of references, in writing conservation laws, and so forth.  Instead, in the geometric approach, helices are treated in exactly the same fashion as straight lines, so stationary helical states and even  their linear stability properties may be considered conveniently.}

\section{Derivation of equations in $SE(3)$ coordinates} \label{derivation-sec}

\revision{Q7}{In this section, we derive the full equations of motion for a self-interacting rod in the discrete and continuous cases.}  This derivation is based on the exact geometric rod theory \cite{SiMaKr1988}, that derives equations equivalent to Kirchhoff's equations for elastic rods using symmetry-reduced variables.  Since the derivation of the exact geometric rod equations for purely elastic rods is well-established, we shall concentrate on the corresponding derivation of the nonlocal equations in the language of \revision{R}{the Lie group $SE(3)$ -- the} Special Euclidean group of orthogonal rotations and translations in $\mathbb{R}^3$.  This familiar Lie group is a semidirect product of $SO(3)$ and $\mathbb{R}^3$ with the following definition of group multiplication: 
\begin{equation}
\left( \Lambda_1, \boldr_1 \right) \cdot \left( \Lambda_2, \boldr_2 \right) = \left( \Lambda_1 \Lambda_2 , \Lambda_1 \boldr_2 + \boldr_1 \right) 
\label{SE3def}
\end{equation} 
for $\Lambda_1, \Lambda_2 \in SO(3)$ and $\boldr_1,\boldr_2\in \mathbb{R}^3$. We use the notation $\mse(3)$ for the corresponding Lie algebra.  \revision{Q1,Q2}{For the reader's convenience, we summarize  the properties of this group, its adjoint and coadjoint actions ${\rm Ad}$, ${\rm Ad}^*$, $ {\rm ad}$, ${\rm ad}^*$ in \ref{sec:appendixA}.  More details can be found in, for example, \cite{Ho2009}.}

\revision{Q1}{
\begin{definition}
A \emph{bouquet} is a rigid, non-deformable assembly of spheres, each characterized by an interaction radius (for short-range interactions like Lennard-Jones), and an electrostatic charge.
\end{definition}
}

\revision{Q1}{A molecule, as considered in this paper, consists of a \revision{R}{rod} that represents the molecule's central axis, with bouquets attached at fixed base points along the \revision{R}{rod}.  Each bouquet is characterized by the spatial position coordinate of its base $\boldr(s)\in\mathbb{R}^3$ and its orientation $\Lambda(s)\in SO(3)$, which together define an element $\sigma(s)=( \Lambda(s), \boldr(s) ) \in SE(3) \cong SO(3)\times \mathbb{R}^3$.}
\revision{Q1}{We are interested in describing the effects of nonlocal interactions, which in this paper denotes all inter-atomic forces that affect parts of the molecule not immediately adjacent to each other. These are all the forces that do not come from elastic deformation, and include, for example, electrostatic and Lennard-Jones forces.}

To describe nonlocal interactions among these bouquets of charges, we define their relative orientation and position variables by the $SE(3)$ product
\begin{equation} 
\eqalign{
\Xi(s,s')=\sigma^{-1} (s) \sigma(s') &= 
\big( 
\Lambda^{-1}(s) \Lambda(s') \, , \, \Lambda^{-1} (s) ( \boldr(s') - \boldr(s) 
\big) \\
&:= 
\big( 
\xi(s,s') \, , \, \bkappa(s,s') 
\big) 
\, , 
}
\label{Xidef} 
\end{equation} 
where $\xi(s,s') \in SO(3)$ and $\bkappa(s,s') \in \mathbb{R}^3$. The position of the $k$-th charge in the bouquet whose base is at $s$ along the center-line of the \revision{R}{rod} is given by
\begin{equation} 
\mathbf{c}_k(s)=\boldr(s) + \Lambda(s) \boldeta_k(s)
\,.
\label{c-def} 
\end{equation}
Here $\boldeta_k(s)$ denotes a vector from the base point at $s$ on the \revision{R}{rod} to the $k$-th off-axis charge of the bouquet as measured in a rigid Cartesian frame with orientation $\Lambda(s)$. Then, the distance $d_{km}(s,s')$ between the $k$-th charge at $s$ and $m$-th charge at $s'$ is given by 
\begin{equation}
\eqalign{
d_{km}(s,s') &= \big| \mathbf{c}_k(s) - \mathbf{c}_m(s') \big|
\\
&= 
\big|
\boldr(s)+ \Lambda(s) \boldeta_k(s) - \boldr(s') - \Lambda(s') \boldeta_m(s') 
\big| 
\nonumber 
\\
&= 
 \big| 
\Lambda^{-1} (s)( \boldr(s) - \boldr(s') ) +\boldeta_k(s) -  \Lambda^{-1}(s)  \Lambda(s') \boldeta_m(s') 
\big| 
\nonumber 
\\ 
&=
\big| 
\bkappa(s,s') +  \boldeta_k(s)  - \xi(s,s')  \boldeta_m(s') 
\big| 
= d_{km} \big(\Xi(s,s') \big) 
\, . 
} 
\end{equation}
\begin{rema} 
In finding particular helical solutions and analyzing their stability, we consider the case of polymers where all the bouquets are exactly the same, so $\boldeta_k(s)$ and $\boldeta_m(s')$ are independent of $s$ and $s'$. Nonetheless, the equations of motion we derive in this section, (\ref{discreteeqs}) and (\ref{continuouseqs}), would also be valid for arbitrary dependence of the bouquet's geometry on position. However, the linear stability analysis performed later will explicitly use the fact that all bouquets are identical, and thus cannot be applied to the conformations of molecules with varying geometry. 
\end{rema} 

Consequently, the total energy of the \revision{R}{rod} due to nonlocal interactions among the charges that compose it is obtained in the continuous case as 
\begin{equation} 
E = \sum_{k,m}\int U
\big( 
d_{km} (\Xi(s,s') )
\big) \, \mbox{d} s \, \mbox{d} s' 
\, , 
\end{equation} 
for interaction potential $U$ between individual pairs of charges.
In the discrete case, the integral becomes a sum and $s$, $s'$ are discrete indices, so that
\begin{equation} 
E = \sum_{s,s',k,m} U
\big( 
d_{km} (\Xi(s,s') )
\big)
\, . 
\end{equation} 
The corresponding contribution of nonlocal interactions to the action in Hamilton's principle for the dynamics of the \revision{R}{rod} is thus (in the discrete case) 
\begin{equation}
\label{Snonlocal}
S_{nl}=-\int \sum_{s,s',k,m} U
\big( 
d_{km} (\Xi(s,s') )
\big) \, \mbox{d} t 
\, . 
\end{equation} 
In order to compute the contribution of the mutual charge interactions to the dynamics of the \revision{R}{rod} one must take the variation of this nonlocal part of the action with respect to the charge conformation $\Xi$. Upon denoting 
\begin{equation} 
\nu(s) = \sigma^{-1}(s) \de \sigma(s) \in \mse(3) \, , 
\end{equation} 
where $\mse(3)$ is the Lie algebra of the Lie group $SE(3)$, we find
\begin{equation}
\eqalign{ 
\de \Xi(s,s') & = \de \big( \sigma^{-1} (s) \sigma(s') \big) = 
- \nu (s) \Xi(s,s') + \Xi(s,s') \nu(s') 
\,.
\label{deXi} 
} 
\end{equation}
\revision{Q2}{We now define a pairing $\big< \cdot, \cdot \big>_{TSE(3)}$ between the elements of $TSE(3)$ and $TSE(3)^*$ -- the tangent and cotangent spaces of $SE(3)$ -- as follows. If a vector $U$ is tangent to $SE(3)$ at point $\sigma \in SE(3)$, and $W$ co-tangent to $SE(3)$ at the same point $\sigma$, then $\sigma^{-1} U$ brings the vector $U$ to the identity element of the group, so $\sigma^{-1} U \in \mse(3)$ is in the Lie algebra. Similarly, $\sigma^{-1} W \in \mse(3)^*$, so a scalar product between these elements can be taken as defined in the appendix. Thus, we define 
\begin{equation} 
\big< U \,, \, W \big>_{TSE(3)} = \big< \sigma^{-1} U \, , \, \sigma^{-1} W \big>_{\mse(3)} \, . 
\label{bracketdef} 
\end{equation} 
In what follows, we drop the subscript $TSE(3)$ from the scalar product as not to burden the notation unnecessarily. 
 }

\revision{Q2, Q3}{In order to use the minimal action principle and derive the equations of motion, we need to take the variations with respect to $\Xi$. The variation $\de \Xi$ defined in (\ref{deXi}) is an element of $TSE(3)$, as $\de \Xi$ is tangent to $SE(3)$ at the point $\Xi$. Then, the derivative  $\de U/ \de \Xi$ is an element of $TSE(3)^*$ as it is co-tangent to the group at the same point $\Xi$. The pairing
 \[ 
 \Big< 
\frac{\de U}{\de \Xi}
\, , \, 
\de \Xi
\Big>
\] 
can be now defined according to (\ref{bracketdef}). 
 }
Consequently, the variation of the nonlocal part of the action in (\ref{Snonlocal}) is given by
\begin{equation}
\eqalign{ 
\de &S_{nl} = -\int \sum_{s,s',k,m}
 \Big< 
\frac{\de U}{\de \Xi} 
\, , \, 
\de \Xi
\Big>\mbox{d} t 
\nonumber 
\\
&=
 -\int \sum_{s,s',k,m} \Big< 
\frac{\de U}{\de \Xi} 
\, , \, 
- \nu (s) \Xi(s,s') + \Xi(s,s') \nu(s') 
\Big>\mbox{d} t 
\nonumber 
\\ 
&=
-\int \sum_{s,s',k,m} \Big< 
-\frac{\de U}{\de \Xi}(s,s') \Xi^{-1}(s,s') + \Xi (s,s') \frac{\de U}{\de \Xi}(s',s)
, \, 
\nu(s) 
\Big> \mbox{d} t 
, 
} 
\end{equation}
in which the last step uses the relation $\Xi(s',s)= \Xi^{-1}(s,s')$ obtained from the definition of $\Xi$ in (\ref{Xidef}). 
\revision{Q8}{We shall now proceed with the computation of the variations of action with respect to all dynamical quantities and thereby obtain the equations of motion.} 
\paragraph{Velocities.}
The local part of the Lagrangian in Hamilton's principle for the dynamics of the \revision{R}{rod} is written by introducing the left-invariant variables 
\begin{equation} 
\mu = \sigma^{-1} \dot{\sigma} = \left(\Lambda \, , \,  \boldr \right)^{-1} \left( \dot{\Lambda} \, , \, \dot{\boldr} \right)
= 
\left( 
\Lambda^{-1} \dot{\Lambda} \, , \, \Lambda^{-1} \dot{\boldr}
\right) \in \mse(3)
\, ,
\label{mudef} 
\end{equation} 
as velocities taking values in the Lie algebra $\mse(3)$.

\paragraph{Elastic deformations.}
In the continuous case, the invariant variables that describe elastic deformations are, 
\begin{equation} 
\lambda= \sigma^{-1} \sigma'  \in \mse(3) 
\,.
\label{lamdefcont} 
\end{equation} 
In the discrete case, following the Moser-Veselov method for numerical discretization of rigid body dynamics \cite{MoVe1991} we set, 
\begin{equation} 
\lambda=\sigma^{-1}(s)\sigma (s+1) \in SE(3) 
\,,
\label{lamdefdisc} 
\end{equation} 
where $s=1,2, \ldots$ is the discrete index labeling a given base.  The elastic part of the Lagrangian will then depend on $\lambda$. 

\paragraph{Compatibility conditions.} 
The \emph{compatibility conditions} are obtained, in the continuous case, by using equality of cross derivatives, so that $\sigma_{st}=\sigma_{ts}$. Differentiating (\ref{mudef}) with respect to $s$ and (\ref{lamdefcont}) with respect to $t$, then subtracting yields
\begin{equation}
\frac{\partial \mu}{\partial s}- \frac{\partial \lambda}{\partial t}=-\lambda \mu + \mu \lambda:= \big[ \mu\, , \, \lambda]_{\mse(3)} 
= {\rm ad}_\mu \lambda \, , 
 \label{compatibilitycont} 
\end{equation} 
where $\big[ \mu\, , \, \lambda]_{\mse(3)}$ is the commutator of $\mu$ and $\lambda$ in $\mse(3)$. In the discrete case, 
we write (\ref{lamdefdisc}) as $\sigma(s+1,t)=\sigma(s,t) \lambda(s,t)$. Differentiating this condition with respect to time, we get 
\begin{equation} 
\frac{\partial \lambda}{\partial t}(s)=\lambda(s) \mu(s+1) - \mu(s) \lambda(s) \, , 
\label{compatibilitydisc1} 
\end{equation} 
or in terms of $\mse(3)$-algebra quantities only 
\begin{equation}
\lambda^{-1} \frac{\partial \lambda}{\partial t}(s)= \mu(s+1) - {\rm Ad} _{\lambda^{-1} }\mu(s)  \, .  
\end{equation}

\paragraph{Dynamical equations: variations}
 In the continuous case, the variations of the velocities on the Lie algebra $\mse(3)$ satisfy,
\begin{equation} 
\de \mu = - \nu \mu + \mu \nu + \frac{\partial \nu}{\partial t} 
\, .
\end{equation} 
In the discrete case, this is replaced by a variation on the Lie group $SE(3)$, 
\begin{equation} 
\lambda^{-1}   \de \lambda(s) =-{\rm Ad}_{\lambda^{-1}} \nu(s) +   \nu(s+1)
\,.
\end{equation} 
For the discrete case, one may then compute the variation of the Lagrangian as,
\begin{equation} 
\eqalign{
& \sum_{s}
 \Big< 
\lambda^{-1}(s)  \frac{\de l}{\de \lambda} (s)
 \, , 
 \lambda^{-1}(s)   \de \lambda (s)
 \Big>
 \\
 & \quad = 
  \sum_{s} 
 \Big< 
\lambda^{-1}(s)  \frac{\de l}{\de \lambda} (s)
 \, , 
-{\rm Ad}_{\lambda^{-1}(s)} \nu(s) +   \nu(s+1)
 \Big>
 \nonumber 
 \\
 & \quad = 
 \sum_{s} 
 \Big< 
-{\rm Ad}^*_{\lambda^{-1}(s)} \Big(  \lambda^{-1}(s)  \frac{\de l}{\de \lambda}(s)  \Big) 
+
\lambda^{-1}(s-1) \frac{\de l}{\de \lambda}(s-1) 
, \,
\nu(s) 
 \Big>.
}
\end{equation} 
In continuous time, the Euler-Poincar\'e equations emerge from the following direct computation as in \cite{HoPu2008,El-etal-2009},
\begin{equation}
\fl
\eqalign{
\int  &
 \int \Big< 
 \frac{\de l}{\de \mu} 
 \, , 
 \de \mu 
 \Big> \mbox{d} t \, \mbox{d} s
 =
 \int \int 
  \Big< 
 \frac{\de l}{\de \mu} 
 \, , \, 
 - \nu \mu + \mu \nu + \frac{\partial \nu}{\partial t}
 \Big>
 \mbox{d} t \, \mbox{d} s
\nonumber 
\\ 
&=
 \int \int 
  \Big< 
 \frac{\de l}{\de \mu} 
 \, , \, 
 {\rm ad}_\mu \nu + \frac{\partial \nu}{\partial t}
 \Big>
 \mbox{d} t \, \mbox{d} s
=
 \int 
\int
 \Big< 
\Big( - \frac{\partial}{\partial t} +{\rm ad}^*_\mu \Big) \frac{\de l}{\de \mu} 
 \, , \, 
 \nu
 \Big>
\mbox{d} t \, \mbox{d} s
\,.
}
\end{equation}
Likewise, the equations of motion in continuous time for the spatially discrete nonlocal \revision{R}{rod}, may be written in $SE(3)$ coordinates as 
\begin{equation} 
\eqalign{
\Big( - \frac{\partial}{\partial t}  +{\rm ad}^*_\mu \Big) & \frac{\de l}{\de \mu} 
-{\rm Ad}_{\lambda^{-1}(s)} \Big( \lambda^{-1}(s) \frac{\de l}{\de \lambda} (s) \Big) + 
  \lambda(s-1) \frac{\de l}{\de \lambda} (s-1)
 \nonumber 
 \\ 
 & -\sum_{s',m,k} 
 -\frac{\de U}{\de \Xi}(s,s') \Xi^{-1}(s,s') + \Xi(s,s') \frac{\de U}{\de \Xi}(s',s)
 =0 
 \, . 
}
\label{discreteeqs}
\end{equation} 
In the continuous case, a similar calculation gives (note that now $\lambda= \sigma^{-1} \sigma' \in \mse(3)$): 
\begin{equation} 
\eqalign{
\Big( - \frac{\partial}{\partial t} & +{\rm ad}^*_\mu \Big) \frac{\de l}{\de \mu} 
+ 
\Big( - \frac{\partial}{\partial s} + {\rm ad}^*_\lambda \Big) \frac{\de l}{\de \lambda} 
 \nonumber 
 \\ 
 & -\sum_{m,k} \int 
 -\frac{\de U}{\de \Xi}(s,s') \Xi^{-1}(s,s') + \Xi(s,s') \frac{\de U}{\de \Xi}(s',s) \mbox{d} s'
 =0 
 \, . 
}
 \label{continuouseqs} 
\end{equation} 
{\bf Remark } 
Notice that (\ref{continuouseqs}) are exactly the equations derived earlier in \cite{HoPu2008,El-etal-2009}. However, the discrete equations (\ref{discreteeqs}) for nonlocally interacting molecules are new, as far as we know.

\begin{lemma} \label{helix-lemma}
For an arbitrary potential $U(d)$, equations (\ref{discreteeqs}) and (\ref{continuouseqs}) reduce to algebraic equations for helical 
solutions. The solutions of these algebraic equations are stationary helical shapes. 
\end{lemma} 
{\bf Proof} Helical configurations in the discrete case are obtained from taking a given element $a \in SE(3)$ and defining $\sigma(s)=a^s$. In the continuous case, a given element $\lambda  \in \mse(3)$ generates $\sigma(s)$ from the differential equation $\sigma'(s)=\sigma(s) \lambda_0$ (exponential map).  In either case, 
\begin{equation} 
\Xi(s,s')=\sigma^{-1} (s) \sigma(s') = \sigma(s'-s) \, . 
\label{Xiss1}
\end{equation}
Then, 
\[ 
d_{km}(s,s')=d_{km}\big(\Xi(s,s') \big) = d_{km}\big( \sigma(s'-s) \big) \, ,
\] 
and therefore 
\[ \frac{\de U}{\de \Xi} (s,s'):= D_1(s'-s), 
\] 
where $D_1$ is some $TSE(3)^*$-valued function that depends only on the difference between $s$ and $s'$. Thus, the right-hand sides under the sum  (\ref{discreteeqs}) and in the integral in  (\ref{continuouseqs}) depend only on the difference 
between $s$ and $s'$, and are therefore \emph{constant}. Likewise, the left-hand sides of these equations are also constant, as $\mu=0$ and $\lambda=a \in SE(3)$ in the discrete case and $\lambda=\sigma_0 \in \mse(3)$ in the continuous case. Thus, integro-differential equations  (\ref{discreteeqs}) and  (\ref{continuouseqs}) reduce to algebraic equations for helical solutions, as long as the potential between two charges depends (in an arbitrary fashion) only on the Euclidean distance between those charges.$\Box$

 Of course, this result had been well-known by molecular biologists for at least 50 years; in their famous paper 
\cite{PaCoBr1951} Pauling, Corey and Branson noted that ``\emph{\ldots It is likely that these [helical] configurations constitute an important part of the structure of both fibrous and globular proteins, as well as synthetic polypeptides}". As it turned out, the helical structures are amazingly robust; they appear for large variety of molecules with widely ranging elastic properties and charge distributions.  Lemma \ref{helix-lemma} provides a mathematical model for this well-established empirical fact. \revision{G}{It is worth noting however that the molecules that are locally helical almost never stay in the shape of a perfect straight helix if they are sufficiently long. This \emph{folding} of molecules is driven by a combination of the elastic and non-local forces and plays a very important role in the ensuing functionality of molecules. The instabilities 
of the helical states will be considered later using geometric approach. }

\begin{rema}
Lemma \ref{helix-lemma} includes, as a particular case, results from previous work on the helical solutions of the Kirchhoff's rod (no nonlocal terms) or uniformly charged Kirchhoff rod \cite{DeLi2009}, with the charges$^1$\footnote[0]{$^1$ Here, we use the word ``charge" somewhat loosely to denote arbitrary potential interaction, like Morse or Lennard-Jones, and not just electrostatic potential.} 
positioned at the rod's axis. 
  In that case, the rigid charge conformations are positioned at $\boldeta(s)=0$, and thus  
\[ 
d(s,s')=\big| \bkappa(s'-s) \big| \, . 
\] 
In this case, the distance $d$ is independent of the mutual orientation $\xi(s,s')$, which is the first part of $SE(3)$ group element $\sigma^{-1} (s)\sigma(s') :=\big(\xi(s,s'), \bkappa(s,s') \big)$. That is, charges arranged on the axis produce no torque. 

With this particular simplification, the computation still proceeds in the same way as before and reduction to algebraic equations still holds. Of course, more work is needed for each particular potential to demonstrate that the algebraic equations actually have a solution, especially if complicated elasticity laws in the \revision{R}{rod} are assumed. Thus, Lemma \ref{helix-lemma} provides a simple justification for helical solutions that have been ubiquitous in the previous literature. However, our result goes further: helical shapes also allow one to search for helical solutions as solutions of algebraic equations for arbitrary charges off axis and for a 
charge distribution at each point of the molecule, possibly generating torque in the molecule.
\end{rema}

\section{Application to a linear rod with naturally straight conformation} \label{linearstrand-sec}
\revision{Q7}{This section demonstrates how to find helical stationary shapes for charge bouquets positioned along  an infinitely long, naturally straight molecule.}

To illustrate the general ideas described above, we shall consider a molecule that consists of a chain of centerline atoms connected with elastic springs, and charges that are attached to each unit of the elastic chain with a rigid charge bouquet.  In the undisturbed (base) configuration, the molecule is straight.  In \revision{R}{what follows, we select a charge bouquet containing two charges of $q_i=\pm 0.3 e$, where $e$ is} the charge of electron. This arrangement approximates any molecule which has a constant dipole moment perpendicular to the axis, for example, vinylidene fluoride oligomers (VDF) \cite{No2003}.  More general charge configurations such as quadrupoles \emph{etc.} can be incorporated by considering more general bouquets. 

\begin{figure}[!ht] 
 \centering
 \includegraphics[width=0.5\textwidth]{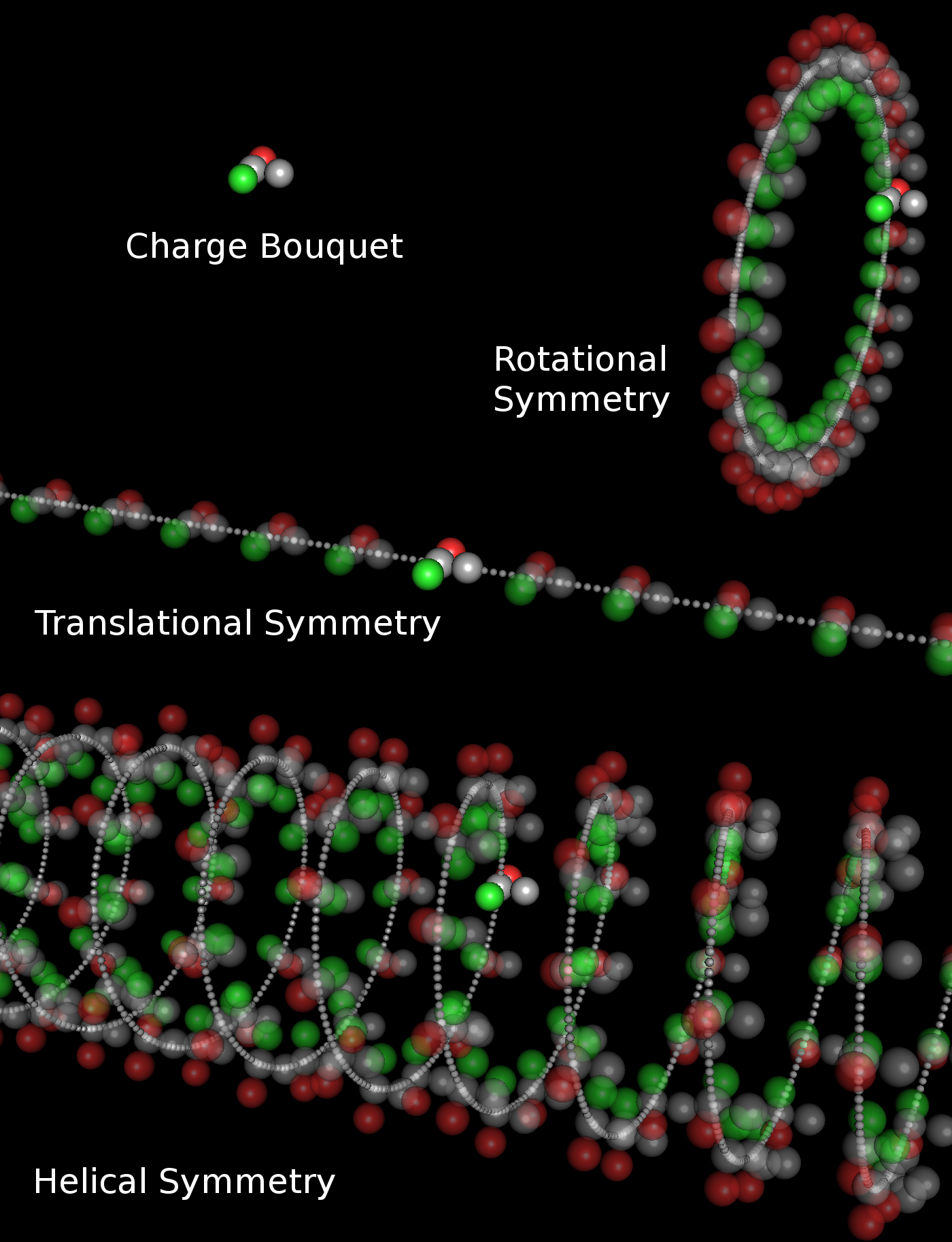}
 \caption{ \label{fig:sketch}\footnotesize
 Top left: an example of a charge bouquet with four atoms. The electrically charged atoms are shown in red and green (positive or negative). Grey atoms are neutral, but still interact with other atoms through Lennard-Jones potential (\protect{\ref{LJ}}). The molecules we consider are long chains of these bouquets. 
 }
 \rem{ 
 Top right: a circular conformation obtained by applying rotational symmetry to the bouquet. Middle: A linear conformation obtained by applying the translational symmetry to the bouquet that are considered to be undisturbed configuration of the molecule. Bottom: A helical conformation obtained by simultaneous application of rotation and translation to the bouquet. 
 } 
\end{figure}

\revision{R}{We assume electrical charges exhibit a screened electrostatic interaction}
\begin{equation} 
E_{ij,C} (s,s')= \frac{q_i(s) \, q_j(s')}{4 \pi \varepsilon_0 d_{ij}(s,s')} e^{-d_{ij}(s,s')/\lambda} \, ,
\label{screenedel}
\end{equation} 
\revision{R}{where $d_{ij}$ is the distance between charges $q_i$ at $s$ and $q_j$ at $s'$, and $\lambda$ denotes the Debye screening length. In what follows, we consider values of $\lambda$ corresponding to various ionic strengths in the solution, from $I=0.001$ M/l (close to de-ionized water) to $I=10$ M/l (an order of magnitude higher than sea water).}  Debye length $\lambda$ and ionic strength $I$ are related by 
\begin{equation}
\lambda=\sqrt{\frac{\epsilon_0 \epsilon_r k_B T}{2 N_A e^2 I}} \, , 
\label{lambdaI}
\end{equation}
where $\epsilon_0$ is the permittivity of space, $\epsilon_r$ is dielectric constant of the media, $k_B$ is Boltzmann's constant and $N_A$ is Avogadro's number. 

We seek stationary states that are invariant with respect to an affine (helical) transformation of space, which rotates a coordinate frame $F$ by a matrix $\Lambda$,\revision{R}{ and performs a translation of an arbitrary vector $\boldr$ as}
\begin{equation} 
(F, \boldr) \rightarrow \left( \Lambda F, \Lambda \boldr + \mathbf{a} \right) \, .
\end{equation} 
\revision{Q1}{In particular, we seek solutions with bouquets spaced uniformly on a helix of radius $R$ and pitch $C$, with the rotation being parallel to the $(x,y)$ plane. By \emph{pitch}, we mean that after making one period of rotation, the atoms are moved by the distance $C$ along the $z$-axis.  The helix is then defined parametrically by} 
 \begin{equation}
  x = R \cos t \, ,  \quad  y = R \sin t \, ,  \quad  z=C t \, ,
  \label{xyzcoordparam}
\end{equation}
or explicitly 
\begin{equation}
  x = R \cos z/C \, ,  \quad  y = R \sin z/C \, ,
\label{xyzcoord}
\end{equation}
where $R \ge 0$ and $C \ne 0$.  The helix is right-handed for $C > 0$ and left-handed for $C < 0$. We use $R$ and $C$ as parameters in our calculations and visualizations.
A continuous helix is a curve that is invariant under a set of transformations consisting of a rotation $\Lambda$ about the $\mathbf{\hat{z}}$ axis and a translation $\mathbf{a}$, parameterized by $z$,
\begin{equation}
 \Lambda(z) = \left[ \begin{array}{c c c}
                \cos \left( \frac{z}{C} \right) & \sin \left( \frac{z}{C} \right) & 0 \\
                -\sin \left( \frac{z}{C} \right) & \cos \left( \frac{z}{C} \right) & 0 \\
                0 & 0 & 1
              \end{array} \right] \, , 
\quad \quad 
  \mathbf{a}(z) = \left[ \begin{array}{c}
                     R \sin \left( \frac{z}{C} \right) \\
                     R \cos \left( \frac{z}{C} \right) \\
                     z
                   \end{array} \right] \, .
                   \label{helicalsymmetry}
\end{equation}
For a discrete helix, we consider values of $t$, (respectively, $z$) that are discrete multiples of some value $T$ (respectively, $z/C$): 
\begin{equation}
  t = n \, T \quad \mbox{resp.} \quad z=n \frac{z}{C} \, , 
  \quad
  \mbox{ where }
  \quad
  n \in \mathbb{Z} \, .
\end{equation}

\begin{figure}[!ht]
 \centering
 \includegraphics[width=0.5\textwidth,angle=90]{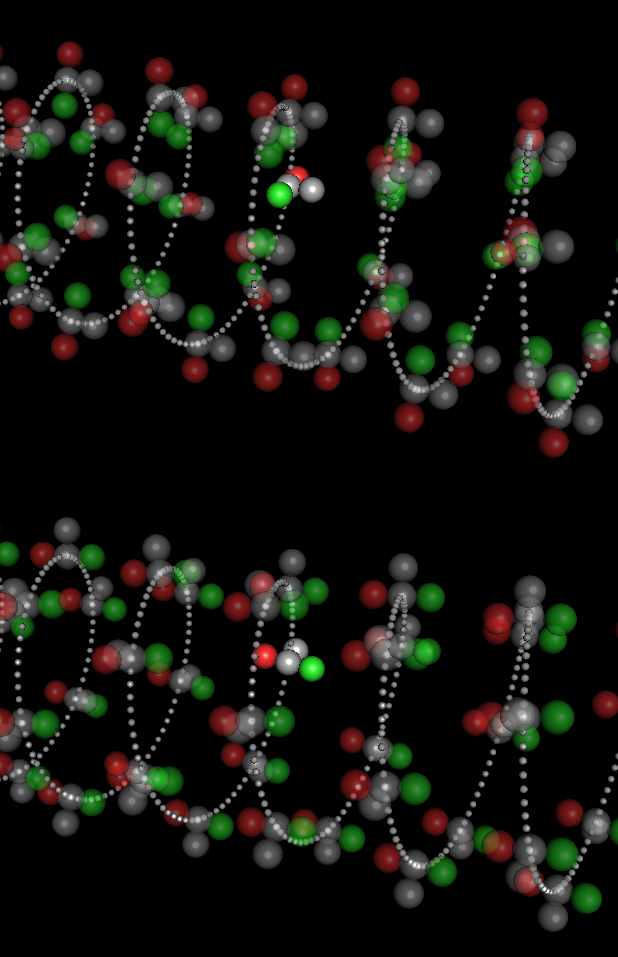}
 \caption{\footnotesize 
 Effect of twist transformation - the twisted bouquet is highlighted.}
 \label{fig:twist}
\end{figure}

\revision{R}{
Bouquets on the helix are oriented such that the helical invariance described above is maintained. Once a single bouquet is specified, at say $z=0$, (\ref{helicalsymmetry}) generates the entire helix structure.  Let $A \in SO(3)$ denote the orientation of the initial bouquet at $z=0$.  In what follows, we limit $A$ to a twist about the tangent to the helix at $z=0$, as shown in Fig.~\ref{fig:twist}, allowing a single angle $\alpha$ to characterize bouquet orientation.}
\begin{equation}
\fl
A(R, C, \alpha) = \left[
\begin{array}{ccc}
  \cos \alpha &
    \frac{-C}{\sqrt{4 \pi^2 R^2 + C^2}} \sin \alpha &
    \frac{2 \pi R}{\sqrt{4 \pi^2 R^2 + C^2}} \sin \alpha \\
  \frac{C}{\sqrt{4 \pi^2 R^2 + C^2}} \sin \alpha &
    \frac{4 \pi^2 R^2 + C^2 \cos \alpha}{4 \pi^2 R^2 + C^2} &
    \frac{2 \pi R C}{4 \pi^2 R^2 + C^2} (1-\cos \alpha) \\
  \frac{-2 \pi R}{\sqrt{4 \pi^2 R^2 + C^2}} \sin \alpha &
    \frac{2 \pi R C}{4 \pi^2 R^2 + C^2} (1-\cos \alpha) &
    \frac{4 \pi^2 R^2 \cos \alpha + C^2}{4 \pi^2 R^2 + C^2} \\
\end{array}
\right] \, .
\end{equation}
The entire helical structure may then be generated by repeated application \revision{R}{}of (\ref{helicalsymmetry}). 

\section{Energy of a helical conformation}

\revision{Q7}{This section explains the interaction energies of a realistic molecule to be used in calculations, both the local (elastic) parts and non-local (electrostatic and Lennard-Jones). } 
\label{sec:minenergy} 
As discussed above, \revision{R}{we investigate the particular example of a molecule of bouquets with two opposite charges $\pm q$ positioned on either side of the polymer axis, each a distance $l$ from that axis.} This simplified model describes a polymer with a constant polarization perpendicular to the polymer's axis. The interaction energies are defined as follows.

\paragraph{Elastic energy.}
\revision{R}{We consider elastic energy as quadratic in bond bend angle,
\begin{equation}
E_{elastic} = \frac{1}{2} \mu \big(\Delta \phi\big)^2 \, , 
\end{equation}
where $\mu$ is a spring constant, and $\Delta \phi$ is the bond bend angle with respect to its unstressed} orientation \emph{in the intrinsic frame}, assumed straight. For naturally helical molecules, $\Delta \phi$ would be the value of rotation angle for \revision{R}{a bouquet} about its rotation axis in the intrinsic frame  from its equilibrium value. In terms of global geometry that would mean that bend energy is punishing creation of local curvature.   A value of $\mu = 3.025 \cdot 10^5 J/(rad^2)Mol = 5.022 \cdot  10^{-19} J/rad^2$ is typical for bend rigidity for a carbon-carbon bond. The bend depends (through a complex algebraic formula that we do not present here) on the radius and pitch \revision{Q1}{(axial distance traversed on each rotation)} of the helix. 

\paragraph{Lennard-Jones energy.}
\revision{R}{To prevent self-intersection, we introduce a truncated Lennard-Jones interaction between charge bouquet nodes with equilibrium distance $d_0$ and potential well depth $\varepsilon$.  When the centers of two bouquet nodes are a distance $d$ from each other, they experience a potential given by}
\begin{equation}
E_{LJ} = \left\{
  \begin{array}{l l}
    \varepsilon \left[
          \left( \frac{d_0}{d} \right)^{12}
      - 2 \left( \frac{d_0}{d} \right)^{6}
      - \frac{1}{3^{12}}
      + \frac{2}{3^6}
    \right] & d < 3 \, d_0 \\
  0 & d \ge 3 \, d_0
  \end{array}
  \right. .
  \label{LJ}
\end{equation}

\paragraph{Charge potential energy.}
As mentioned when the model was introduced, each bouquet \revision{R}{has} charges of $q_{1,2}(s) = \pm 0.3 e$ that interact with each other through the screened Coulomb potential (\ref{screenedel}) with Debye length $\lambda$. 

\paragraph{Total energy of a given conformation.}
\revision{R}{Given a bouquet configuration and values for $R$, $C$, and $\alpha$, we compute the total energy by choosing a reference bouquet, then working outward along the helix in both directions by helical symmetry.  For each bouquet we reach, its contribution is the sum of the Lennard-Jones and Coulomb interactions of the bouquet nodes with the reference bouquet nodes.  The sum of these, plus the elastic energy corresponding to the given $R$ and $C$ give the energy per bouquet (energy density) of the given conformation.}

\begin{figure}[!ht]
 \centering 
 \includegraphics[height=2.8in]{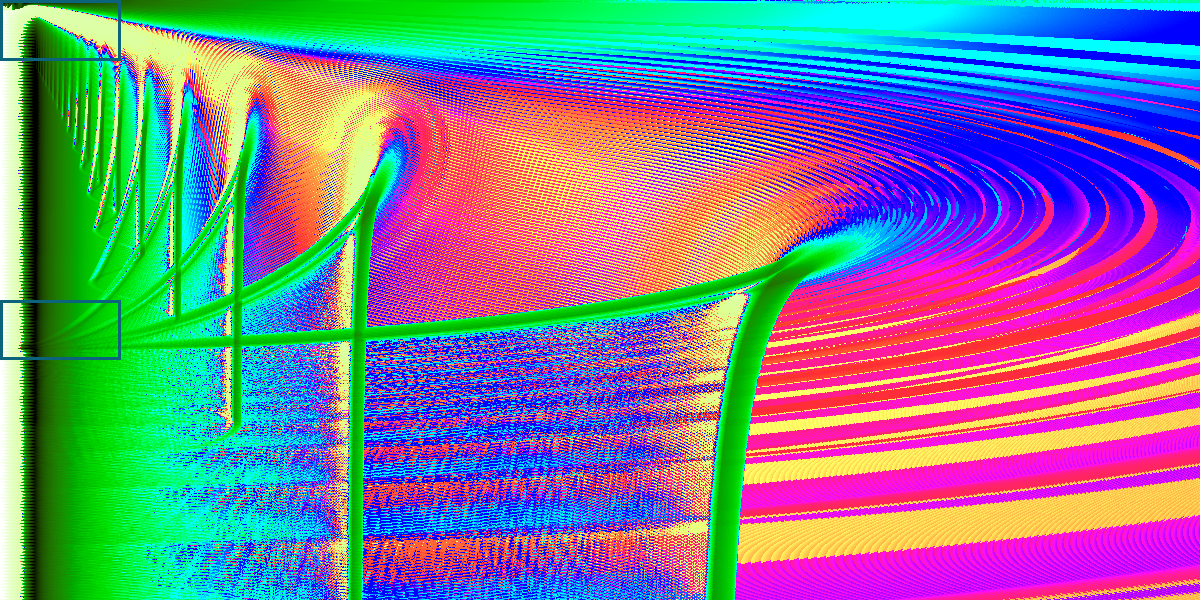}
 \includegraphics[height=2.8in]{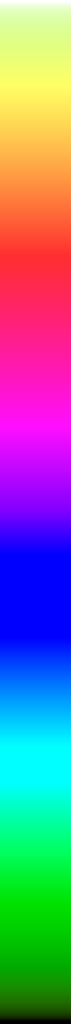}
 \caption{\footnotesize The energy landscape for a given ionic strength $I=0.001 M/l$. The vertical coordinate is radius $R$ and the horizontal coordinate is pitch $C$. Energy is also dependent on the rotation of the bouquet around the axis $\alpha$, but this coordinate has been projected out onto two dimensions by taking a minimum over all rotation angles for the bouquet. That projection preserves the minima of the energy. Notice that this energy plot provides a complete analysis for all helical conformations. Helical conformations that are potential minima are concentrated on the left-hand side (small pitch, increasing radius) and upper left-hand corner (small radius, increasing pitch. Below, Figure~\protect{\ref{fig:blow-up}} shows blow-ups of the framed rectangular regions. 
 }
 \label{fig:energylandscape}
\end{figure}

As an example, we demonstrate the particular energy landscape for $I=0.001 M/l$ in Fig.~\ref{fig:energylandscape}. The vertical \revision{R}{axis} is helix radius $R$, the horizontal \revision{R}{axis} is helix pitch $C$, with $(R,C)=(0,0)$ in the upper left corner. The energy scale bar is shown on the right, with lowest energies in black, highest in white. The left \revision{R}{edge} and the upper left-hand corner contain most of the energy minima. The minima along the left-hand \revision{R}{edge}, corresponding to small pitch $C$ (on the order of the size of the bouquet), corresponds to the helices solutions that are being ``ratcheted" into tighter and tighter conformations as we proceed upward (smaller $R$). The values of the energy in this graph are minima over all twists \revision{R}{(values of $\alpha$)} of the bouquet, as shown in Fig.~\ref{fig:sketch}. The local minima of this energy landscape are stable helical conformations.

\begin{figure}[!ht]
 \centering 
\includegraphics[width=0.8\textwidth]{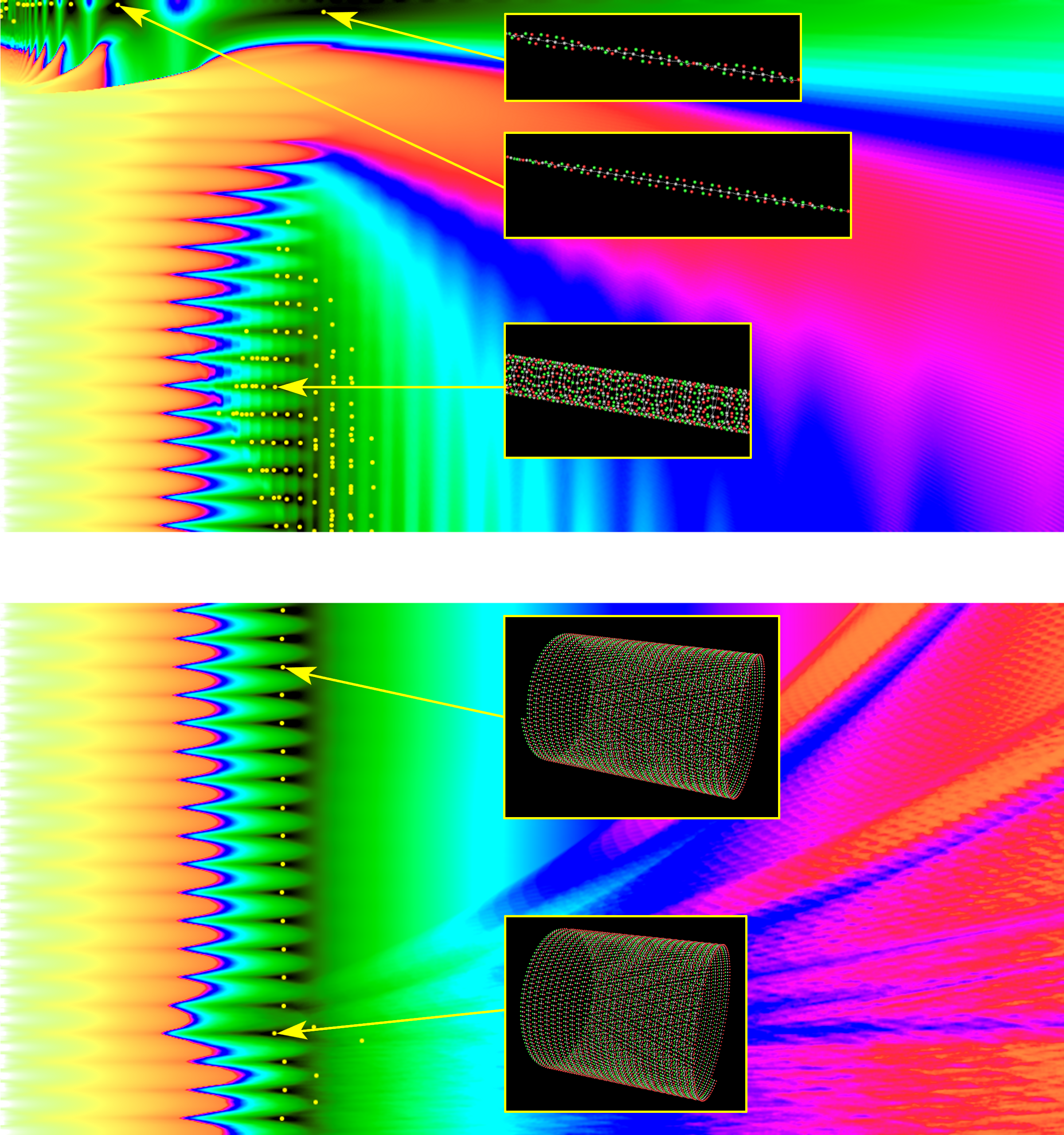}
 \includegraphics[height=0.4\textwidth]{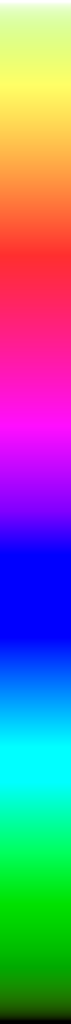}
 \caption{\footnotesize
 Blow-up of energy landscape for the boxes in Fig.~\protect{\ref{fig:energylandscape}}.  The energy scale has been adjusted to show more detail, and is shown on the right with lowest energies in black, highest energies in white. The vertical coordinate is radius $R$ and horizontal coordinate is pitch $C$. Top: upper left corner box in Fig.~\ref{fig:energylandscape}. Bottom: box in the middle of left edge in Fig.~\ref{fig:energylandscape}. Several examples of the helical conformations corresponding to different energy minima are presented as inserts with yellow arrows indicating the corresponding energy minima.
   }
 \label{fig:blow-up}
\end{figure}

In order to further elucidate the structure of the energy landscape, Fig.~\ref{fig:blow-up} presents a blow-up of the two boxes in Fig.~\ref{fig:energylandscape}. Yellow dots correspond to energy minima and thus show stationary conformations.  For completeness, in \ref{sec:A-supp}, we show the change of energy landscape as the ionic strength of the surrounding solution varies from $0.001$ M/l to $0.016$ M/l.  Since we assume that the base conformation of our molecule is linear, a weakly ionized solution leads to longer electrostatic interaction, resulting in richer structure.  Helical conformations become increasingly unlikely for strongly ionized solutions that make long-range electrostatic forces weaker. For strongly ionized solutions, elastic forces tend to unwind the helices into straight lines.

As shown in Fig.~\ref{fig:blow-up}, even this simple molecule, when deformed into helical configurations, shows quite a complex and intriguing energy landscape. While the details of the landscape depend on the individual molecular parameters, the presence of ``ratcheting" helical states is, we believe, typical for all molecules of this type. In this figure, several examples of the helical conformations corresponding to different energy minima are presented as inserts.  There are two types of helical conformations. The ``twist" type is obtained by twisting a base straight conformation, and is characterized by small radius $R$ and increasing pitch $C$. In a given energy landscape, there are only finitely many conformations of this type, as increasing the twist beyond certain level will cause elastic forces that are too large to be balanced by  electrostatic attraction. We show two examples of such twist conformations in the top part of this figure. Another type of helical solution is given by the ratcheting states, characterized by a pitch on the order of bouquet's size and increasing radius. This  is shown both in the top and bottom parts of the figure. There are infinitely many of these ratcheting states, as the radius can increase (in principle) to arbitrarily large values. Energy minima then occur every time the opposite charges line up on sequential rolls of the helix.  That explains the regularity of occurrence of those minima. They occur every time the helix circumference increases by the distance separating the bouquets, $l$. This leads to an increase of radius between neighboring ratcheting conformations as  $\Delta R= l /(2 \pi)$. 

To illustrate how varying ionic strength of the solution changes the energy landscape of helical conformations, we compute 10 energy landscapes varying from $I=0.001 M/l$ to $I=0.16 M/l$. These energy landscapes are presented, in an animated fashion and with the same orientation and axes as in Fig.~\ref{fig:energylandscape}, in the supplement. 
From this, it is clear that the variation of ionic strength and hence Debye length, has a profound effect on the energy landscape. As we see below, the ionic strength is also of particular importance for linear stability.

\rem{ 
\begin{figure}[!ht]
 \centering 
 \caption{(Supplementary Figure) Energy landscapes for different values of ionic strengths $I$, expressed in Moles/l. A value $I=0.001 M/l$ is a very weakly ionized solution, whereas $I=0.16 M/l$ is a rather strongly ionized solution.   
 }
 \label{fig:landscapes}
\end{figure}
}

\section{Multiple helical conformations}\label{multi-helices-sec}

\revision{Q7}{In this section, we explore generalizations of the helical shapes considered above.}
We show how to compute a particular molecular shape \revision{R}{that we refer to as} a \emph{2-helix} (and, in general, $n$-helix).  By 2-helix we mean a conformation where \revision{Q9}{a group consisting of two bouquets with some relative position and orientation is repeated along a helix.  Alternatively, a 2-helix can be seen as two helices of single bouquets arranged so the polymer axis passes through bouquets from alternating helices over its length.}   The relative orientation of bouquets in the group can be arbitrary. Such structures have been suggested and analyzed in \cite{Qu1999} in the context of describing given molecular configurations: $\beta$-helix structure (2-helix) or collagen (3-helix). These structures may be obtained as natural energy minima for a given molecule. 
In particular, we present an example of an equilibrium configuration of a 2-helix using the bouquet considered in the previous section. An extension of these ideas is possible for \revision{R}{$n$-helices with $n > 2$}.  In general, $n$-helices are combinations of $n$ helical conformations, with $2$, $3$, \ldots, $n$ being obtained from the first one by a shift, rotation and a twist of the base bouquet. A detailed map of the energy landscape for an $n$-helix is problematic even for moderate $n$, because the number of parameters involved in defining the conformation of such a shape may increase prohibitively with $n$, as the position and orientation of each bouquet is required in defining a repeated group in an $n$-helix.

\begin{figure}[!ht]
 \centering 
\includegraphics[width=0.51\textwidth]{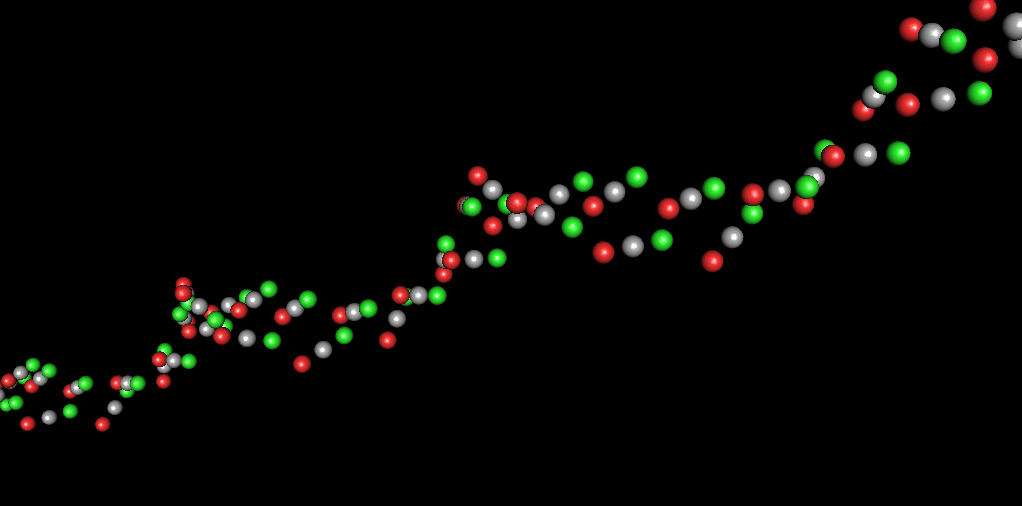}
\includegraphics[width=0.395\textwidth]{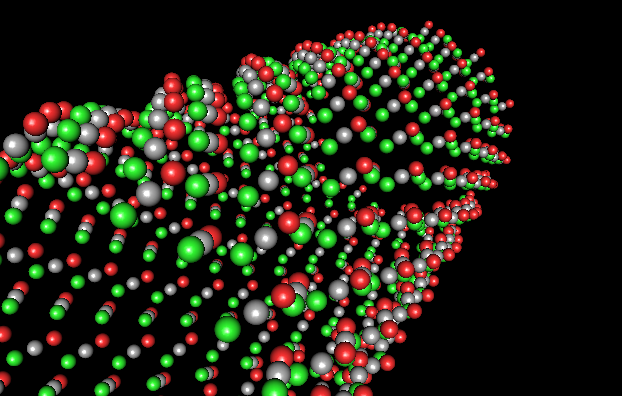}
 \caption{\footnotesize
 2-helix conformations for the ionic strength I$ = 0.001$ M/l, and red/green charges in each bouquet of $\pm 0.8 e$. The elastic centerline (not shown) goes through gray spheres that are the centers of the bouquets. The conformation is not a perfect helix, but instead consists of two helices, in which the second helix has undergone a twist, shift and rotation with respect to the first one. The distance between the bases of the bouquets (gray spheres) is 2 \AA.
   }
 \label{fig:2-helix}
\end{figure}

For a 2-helix, we optimize the energy over both the shape of the helix and the relative configuration of two bouquets. A minimal energy over these variables will, by the helical symmetry, provide an exact stationary conformation for the whole molecule. Two examples of such conformations are given in Fig. \ref{fig:2-helix}.  Details of calculations of more general multi-helices will be discussed elsewhere. 

\section{Linear stability analysis} \label{linstability-sec}
\revision{Q7}{ 
This section derives the dispersion relations for the linear stability of helical polymers, based on the linearizations of (\ref{discreteeqs}) and (\ref{continuouseqs}) about a helical state that is assumed to be a stationary solution. As we show in this section, the use of geometric methods yields an exact dispersion relation.} 

The energy landscape for charged polymers with nonlocal interactions is complex, even when limited to helical shapes, because of the large number of energy minima in that ``helical universe". It is natural to pose the question of linear stability and selection of the stationary shapes. Here, the geometric approach allows us to find the exact dispersion relation for the stability of the helical shapes. As far as we know, no such work has been undertaken before, perhaps because of the complexity of the linear stability analysis when using traditional methods. 

By using an $SE(3)$ method that naturally accounts for the helical geometry, one may investigate the linear stability of a general helical polymer with torque-inducing non-local self-interactions. Studies of the stability of helical elastic \revision{R}{rods} have been undertaken before, with most studies concentrating on the linearization of the traditional Kirchhoff equations about the stationary helical states \cite{GoTa1996,GoTa1997,GoTa2000,ShGo2000,GoNiTa2001,LaGoTa2005}. The focus of these works was on the stability analysis based on increasingly complex elastic properties of the rod. Alternatively, the work \cite{BiCoZh2004} investigated the stability of elastic rods using the exact geometric rod theory and applied it to DNA dynamics. All these works have used the continuous model of elastic rods. Our results differ in two ways from previous studies. First, our results are formulated for spatially discrete rods, as discussed in the previous sections of this paper. Second, and more importantly, our stability analysis includes non-local interaction of charges that in general occupy positions \emph{off the axis} of the elastic rod. The forces on these charges generate torques acting on the rod's centerline. These torques are absent when one considers purely elastic \revision{R}{rods}, or when the charges are distributed only along the centerline. 

This section shows that the presence of the torques due to non-local interaction of off-axis charges generates an \emph{instability} of the \revision{R}{rod} and that the instability appears even for the simplest possible states -- the linear \revision{R}{rod}. The instability due to nonlocal torques is new, as far as we are aware. 

Obtaining similar results using Kirchhoff's rod equations would be problematic. The main difficulty consists in finding the Euclidean distance between two arbitrary points on a \revision{R}{rod}'s axis using coordinates intrinsic to the \revision{R}{rod}. In addition, the present computation shows the geometric origin of the exact dispersion relation for arbitrary helical configuration of a \revision{R}{rod} (without charges), something that was noticed already in \cite{GoTa1996}. In our opinion, the present method based on exact geometric theory  is more straightforward,  algorithmic and compact than the corresponding linear analysis of Kirchhoff equations.

Suppose we have a helical configuration arising from successive repetition of a given element 
$a\in SE(3)$, so that $\sigma(s)=a^s$ and $\Xi(s,s')=a^{s'-s}$. Let us linearize about this solution, so 
\begin{equation} 
\sigma(s) = a^s + \epsilon a^s \psi_1(s,t) +O(\epsilon^2) \, . 
\label{lin0} 
\end{equation}
The linearization written symbolically as $\sigma=\sigma_0+\epsilon \sigma_1$ is defined by 
\begin{equation} 
\psi_1=\sigma_0^{-s} \sigma_1(s):=a^{-s} \sigma_1 (s) \in \mse(3)
\,.
\label{lin00} 
\end{equation}
As it turns out, this substitution leads to an exact dispersion relation for $\sigma_1$. 
\revision{Q8}{ We define the $O(\epsilon)$ perturbations in velocity and deformation as follows. }
First, perturbations of the velocity are determined, as
\begin{equation} 
\mu(s,t) = \epsilon \mu_1 (s,t) + O(\epsilon^2) 
\, , \quad 
\mu_1 = \sigma_0 ^{-1} \dot \sigma_1 = \dot \psi_1 \in \mse(3) \, . 
\end{equation} 
In the discrete case, 
\begin{equation} 
\lambda(s,t) =a + \epsilon \lambda_1 + O(\epsilon^2) 
\, ,
\end{equation} 
where
\begin{equation} 
\lambda_1=- \sigma_0^{-1}(s) \sigma_1(s) a + \sigma_0^{-1} (s) \sigma_1(s+1) = -\psi_1(s) a+ \psi_1(s+1) \in \mse(3) \, . 
\end{equation}
In the corresponding continuous case, one denotes  $\lambda_0= \sigma^{-1} (s) \sigma'(s) = \Gamma$ and finds, 
\begin{equation}
\eqalign{
\lambda(s,t) &= \Gamma +\epsilon \lambda_1 + O(\epsilon^2)\\
&\lambda_1 = - \sigma_0^{-1} \sigma_1 \Gamma  + \sigma_0 ^{-1}\sigma_1'(s) = - \psi_1 \Gamma+ \psi_1'  \in \mse(3)\, . 
}
\end{equation} 
For simplicity (in order to keep the formulas compact), we shall assume for the discrete case that 
\begin{equation} 
 \frac{\de l}{\de \mu} = \Pi_0 + \epsilon I \mu_1  + \ldots \, , \quad  
\lambda^{-1} \frac{\de l}{\de \lambda} = K_0 + \epsilon  K_1(s) + \ldots  \, . 
\end{equation} 
\revision{Q8}{Next, we compute the linearization of the nonlinear terms due to elasticity.} 
In order to find the linearization of the sum of terms
\begin{equation}
{\rm Ad}^*_{\lambda^{-1}(s)} 
\Big( 
\lambda^{-1}(s) \frac{\de l}{\de \lambda} (s) 
\Big) 
+ \lambda^{-1}(s-1) \frac{\de l}{\de \lambda} (s-1) \, ,
\label{lambdaterms} 
\end{equation} 
in equation (\ref{discreteeqs}) we utilize the following proposition.
\begin{proposition} 
Suppose $G$ is a Lie group with Lie algebra $\mathfrak{g}$ and $\langle \cdot \,,\, \cdot \rangle:  \mathfrak{g}^* \times \mathfrak{g} \to \mathbb{R}$ is a pairing between the Lie algebra and its dual. 
Suppose $A(\epsilon) \in G$ is a curve in $G$ with  $A(0)=A_0$ and $A_0^{-1} A'(0)=a \in \mathfrak{g}$ ,  $\alpha(\epsilon) \in \mathfrak{g}^*$, and 
$\alpha'(0)=\xi$. Then (see, for example, \cite{Ho2009}, p.60)  
\begin{equation} 
\frac{\partial }{\partial \epsilon} {\rm Ad}^*_{A^{-1}(\epsilon)} \alpha(\epsilon) \Bigg|_{\epsilon=0}  
= 
{\rm Ad}^*_{A_0^{-1}} \Big(  \xi - {\rm ad^*} _a \alpha_0 \Big) \, . 
\label{adstarderiv}
\end{equation} 
\end{proposition}
Note that $\lambda^{-1} \de l/ \de \lambda$ has the physical meaning of the local stress in the body coordinate frame at point $s$. 

Then, writing 
\[ 
\lambda^{-1} \frac{\de l}{\de \lambda}(s)=K_0 + \epsilon K_1(s)+ O(\epsilon^2) \, , 
\] 
where $K_0 \,, K_1  \in \mse(3)^*$.  
Hence,   the linearization of (\ref{lambdaterms})  is computed as follows 
\begin{eqnarray} 
&&\frac{\partial }{\partial \epsilon} \Bigg|_{\epsilon=0} \Bigg( 
-{\rm Ad}^*_{\lambda^{-1}(s)} 
\Big( 
\lambda^{-1}(s) \frac{\de l}{\de \lambda} (s) 
\Big) 
+ \lambda^{-1}(s-1) \frac{\de l}{\de \lambda} (s-1)
\Bigg)
\nonumber\\
&&\hspace{1cm}
= 
- {\rm Ad}^*_{\lambda_0^{-1}} \Big( K_0 - {\rm ad}^*_{\psi_1} K_1(s) \Big) (s) + K_1(s-1)\, , 
\label{lambdaterms2} 
\end{eqnarray} 
where $\psi_1(s):=\lambda_0^{-1} (s)\lambda_1(s) =  \in \mse(3)$.

The linearization of the nonlocal terms in equation (\ref{discreteeqs}) is less straightforward 
and will be outlined in its own section below. For now, we assume it is possible to compute that linearization, and it is 
described by some linear operator $\mathbb{L}(\Xi (s,s')) \psi_1(s)$, which is defined as follows. Consider an arbitrary 
$\eta \in \mse(3)$ and define the \emph{scalar} function of $s$ by the following pairing,
\begin{equation} 
I (s)=
\int \big< 
-\frac{\de U}{\de \Xi}(s,s') \Xi(s,s') + \Xi^{-1} (s,s') \frac{\de U}{\de \Xi}(s',s)
\, , \, 
\eta 
\big> \mbox{d} s'
\,.
\label{Idef}
\end{equation} 
The nonlocal term takes values in the space $\mse(3)^*$, so the pairing in (\ref{Idef}) indeed defines a scalar function. We need to find its linearization with respect to $\psi=\sigma^{-1} \de \sigma \in \mse(3)$. For this, we compute the derivative of $I(s)$ with respect to $\Xi$ according to 
\begin{equation}
\de I = 
\Big<  
\frac{\de I (\Xi,\eta)}{\de \Xi} \, , \,  \Xi_1
\Big> = 
\Big<  
\Xi^{-1} \frac{\de I(\Xi,\eta)}{\de \Xi} \, , \,  \Xi^{-1} \Xi_1
\Big> \, , 
\label{delcalc}
\end{equation} 
where $\psi_1 = \Xi^{-1} \Xi_1 \in \mse(3)$ is the linearization with respect to $\Xi$. To complete this calculation, we need 
to express $\Xi^{-1} \Xi_1$ in terms of $\psi_1$. This step proceeds as follows. The linearization of $\Xi(s,s')$ in (\ref{Xiss1}) gives 
\begin{equation} 
\Xi(s,s') = a^{s'-s} + \epsilon \Xi_1 (s,s') + O(\epsilon^2) \, , 
\end{equation} 
where 
\begin{eqnarray} 
\Xi_1(s,s') &=& - a^{-s} \sigma_1(s) a^{s'-s} + a^{-s} \sigma_1(s') 
\nonumber\\
&=& - \psi_1(s) a^{s'-s}+a^{s'-s} \psi_1 (s') 
\nonumber\\
&=& a^{s'-s} \big( \psi_1(s') -{\rm Ad}_{a^{s-s'} } \psi_1(s) \big) 
\, .
\end{eqnarray} 
Consequently, the quantity $\Xi^{-1} \Xi_1$ is given by
\begin{equation} 
\Xi^{-1} \Xi_1 = -{\rm Ad}_{a^{s-s'} } \psi_1(s) +\psi_1 (s') 
\, , 
\end{equation} 
and we have 
\begin{equation}
\fl 
\de I
= 
\Big< 
- {\rm Ad}^*_{a^{s-s'} }  \Big( \Xi^{-1} (s,s') \frac{\de I(\Xi(s,s') ,\eta)}{\de \Xi} \Big) 
+ \Xi (s',s) \frac{\de I(\Xi(s',s) ,\eta)}{\de \Xi}  \, , \, 
\psi_1
\Big>.
\label{dei2}
\end{equation} 
Finally, since $I(s)$ in (\ref{Idef}) is a linear function of an arbitrary $\eta$, re-arranging expression (\ref{dei2}) into a scalar product of an $\mse(3)^*$-valued function with $\eta$ will give the desired linearization operator $\mathbb{L}(\Xi (s,s')) \psi_1(s)$, from 
\begin{equation} 
\de I
=: 
\big< 
\mathbb{L}(\Xi (s,s')) \psi_1(s) \, , \, \eta 
\big> .
\label{dei02}
\end{equation}

The equations simplify further upon noticing that for a stationary helical solution, $\sigma=\sigma_0 a^s$ where $a \in SE(3)$ is a given element, so the expression $\lambda_0=\sigma_0^{-1} (s) \sigma(s+1)=a$ is independent of $s$. 
Then, the linearization of equation (\ref{discreteeqs}) in the discrete case is 
\begin{equation}
\eqalign{
 - \frac{\partial^2}{\partial t^2}   I  \psi_1 
  & - {\rm Ad}^*_{a^{-1}} \Big( K_1(s)  - {\rm ad}^*_{\psi_1 }(s) K_0 \Big) (s) + K_1 (s-1) \\
 &= 
 \sum_{s',m,k} 
\mathbb{L}\big(\Xi (s,s')\big) \psi_1(s)
 \, . 
}
\end{equation} 
It is natural to posit the following ansatz: 
\begin{equation} 
\fl
K_1(s)=a^{-s} 
\Big[ J \big( \psi_1(s+1)- \psi_1(s) \big) \Big] a^{s} 
:={\rm Ad}^*_{a^s}  \Big( J \big( \psi_1(s+1)- \psi_1(s) \big)  \Big) \, , 
\label{K1ansatz}
\end{equation}
where $\psi_1(s) \in \mse(3)$ and $J: \mse(3) \rightarrow \mse(3)^*$ is a linear operator having the physical meaning of the rigidity matrix. Notice that the linearized system of coordinates is written at the point $s$, but it encounters the value of the stress at the point $s-1$. In order to connect this stress with the coordinate system at the point $s$, we will need to transform the coordinates to $s-1$, by shifting one step forward on the helix.  We thus need to compute ${\rm Ad}^*_{a^{-1}}$ of the term 
evaluated at $s-1$, \emph{i.e.} 
\[ 
K_1(s-1)={\rm Ad^*}_{a^{-1} } \Big[ J \big( \psi_1 (s) - \psi_1(s-1) \big) \Big]  .
\] 
%
%
Then, the linearization of the discrete case gives 
\begin{equation}
\eqalign{
 - \frac{\partial^2}{\partial t^2}   I  \psi_1 
   - {\rm Ad}^*_{a^{-1}} \Big[ J \Big( &\psi_1(s+1)- 2 \psi_1(s) + \psi_1(s-1) \Big) \Big]
 \nonumber 
 \\
 &+ {\rm ad}^*_{\psi_1(s) } K_0 =
 \sum_{s',m,k} 
\mathbb{L}\big(\Xi (s,s')\big) \psi_1(s)
 \, . 
  \label{disp0d} 
}
\end{equation} 
In the continuous case, the corresponding linearization of equation (\ref{continuouseqs}) gives 
\begin{equation} 
\eqalign{
 - \frac{\partial^2}{\partial t^2}   I  \psi_1
 +  & \Big(-\frac{\partial}{\partial s} + {\rm ad}^*_\Gamma \Big) J \psi_1' + {\rm ad}^*_{\psi_1}  \Gamma
 = 
 \sum_{m,k} \int
\mathbb{L}\big(\Xi (s,s')\big) \psi_1(s) 
 \mbox{d} s' 
 \, . 
}
\label{disp0c} 
\end{equation} 
Here, again $\psi = \lambda_0^{-1}(s) \lambda_1(s) \in \mse(3)$. 

Further simplification can be obtained for the nonlocal term for the stationary helical state $\sigma(s)=\sigma_0 a^s$. The invariant variable $\Xi=\sigma^{-1}(s) \sigma (s')=a^{s-s'}$ depends only on the difference between $s$ and $s'$. Thus, all of the derivatives of the potential energy with respect to $\Xi$ when evaluated at the helical configuration depend only on the difference between $s$ and $s'$. In other words, we have 
\begin{equation}
\eqalign{
&\Xi_0^{-1} \frac{\de U}{\de \Xi}_0(s,s') = D_1(s-s')  \in \mse(3) ^* \, , \\
&\Xi_0^{-1} \frac{\de }{\de \Xi}_0 \Big( \Xi_0^{-1}  \frac{\de^2 U}{\de \Xi}_0(s,s') \Big)  = D_2(s-s') 
\, ,
}
\end{equation} 
with $D_2 \alpha \in \mse(3)^*$ for any $\alpha \in \mse(3)$, so $D_2: \mse(3) \rightarrow \mse(3)^*$. 

Since all the functions on the right-hand side depend only on the difference 
$s-s'$, the integrals or sums become convolution integrals. Fourier transforming then allows the exact dispersion relation to be obtained, as follows. Let us consider 
\begin{equation} 
\psi_1(s,t)=  S  e^{-i \omega t + i k  s} \, , \quad S \in \mse(3) \, . 
\end{equation} 
Here $s$ is an integer and $k$ is the dimensionless wave number, measured in the units of $2 \pi/l_0$, where $l_0$ is the distance between the elements of the helical chain. 
Consequently, assuming $S$ is real, the linearized equation (\ref{disp0d}) gives the following dispersion relation: 
\begin{equation}
\eqalign{
 \omega^2  I S     
   - 4 \big(  \sin^{2} \frac{k}{2}  \big) {\rm Ad}^*_{a^{-1}} \Big( J S - {\rm ad}^*_S K_0 \Big) 
 = 
 \sum_{s',m,k} 
 \mathbb{L} \big(s' \big) S \, .
}
\label{disp0}
\end{equation} 
 Note that in the absence of non-local interactions, the basic helix must be unstressed, $K_0=0$, so $\omega^2= \lambda$ are given by the generalized eigenvalues of the problem 
\begin{equation} 
  4 \sin^{2} \frac{k}{2} {\rm Ad}^*_{a^{-1}} \Big( J S \Big)  =   I \lambda S \, .
\end{equation} 
 From physical principles, we require all the generalized eigenvalues of the matrices $J$ and $I$ to satisfy $\lambda= \omega^2>0$, so that all purely elastic helices in stationary conformations are neutrally stable. The spatially discrete dispersion relation (\ref{disp0}) converges to the dispersion relation for the continuum case in the limit $k \rightarrow 0$, $a \rightarrow {\rm Id}_{SE(3)}$, which is 
\begin{equation} 
  k^2 {\rm Ad}^*_{a^{-1}} \Big( J S \Big)  =   I \lambda S \, .  
\end{equation} 
 
Again, the right-hand side of (\ref{disp0}) is a function of $s'-s$ only, while the left-hand side is a constant. Upon summation over $s'$, the dependence on $s$ disappears and the dispersion relation is obtained by setting $s=0$ on the right-hand side:
\begin{equation}
\eqalign{
 \omega^2 I S
   - \big( \sin^{2} \frac{k}{2} \big) {\rm Ad}^*_{a^{-1}} \Big( J S - {\rm ad}^*_S K_0 \Big) 
 = 
 \sum_{s',m,k} 
 \mathbb{L} \big(s' \big) S 
 \, .
}
\label{dispfin} 
\end{equation}
The right-hand side is a linear operator acting on $S$. Instability corresponds to generalized eigenvalues $\omega$ of equation (\ref{dispfin})  having a positive imaginary part. As it turns out, all eigenvalues $\lambda=\omega^2$ are real, so it 
is enough to identify the case $\lambda<0$ as instability. However, the consideration of discrete \revision{R}{rods} puts an interesting spin on this problem that we will consider below. 
 
 \section{Computation of derivatives for potential energy} \label{comp-derivs-sec}
Equation (\ref{dispfin}) provides the stability analysis for an arbitrary interaction potential $U(d)$. However, for particular applications it is important to explicitly compute the linearization operators in (\ref{dispfin}). Again, geometric methods will be advantageous. 
 \revision{Q7}{ 
 Therefore, this section computes the linearization of the nonlocal terms for the dispersion relation. We believe it is advantageous to show this computation in some detail, as it is not trivial. 
 }

We shall perform the computation only for the discrete case. The continuous case is derived similarly with the change of sums with respect to $s$ and $s'$ into integrals where necessary. 
\revision{Q8}{One has to be careful here, as we need to take derivatives of quantities that take values in $T_eSE(3)$ and $T_eSE(3)^*$. The most straightforward way, least likely to lead to a mistake, is to define a corresponding  scalar functional by bringing these quantities to the Lie algebra and then pairing them with the corresponding \emph{fixed} element from the dual. 
The derivatives  will then be given by whatever term is paired the chosen fixed element. This is akin to the weak computations of functional derivatives, only performed  with geometric quantities. 
} 
First, notice that 
\begin{equation} 
 \frac{\de U}{\de \Xi} (s,s') \Xi^{-1}(s,s') = {\rm Ad}^*_{\Xi^{-1}(s,s')} \Big(   \Xi^{-1} (s,s') \frac{\de U}{\de \Xi} (s,s') \Big) \, ,
 \label{term1}
\end{equation} 
and 
\begin{equation} 
\Xi(s,s') \frac{\de U}{\de \Xi} (s',s) = \Big(   \Xi^{-1} (s,s') \frac{\de U}{\de \Xi} (s,s') \Big)  \Bigg|_{s \leftrightarrow s'} 
\,.
\label{term2}
\end{equation} 
Thus, we start the computation of the linearization operator $D_2$ for the nonlocal term by linearizing the expression 
$\Xi^{-1} \frac{\de U}{\de \Xi} \in \mse(3)^*$. Let us consider $\Xi=(\xi, \bkappa)$ with 
$\Xi^{-1}=(\xi^{-1}, - \xi^{-1} \bkappa)$, an arbitrary element $(\boldphi, \boldpsi) \in \mse(3)$ and a scalar functional 
\begin{equation} 
I_1= 
\Big< 
\Xi^{-1} \frac{\de U(d_{km})}{\de \Xi}
\, , \, 
(\bmu, \boldalpha)
\Big>.
\label{I1def} 
\end{equation}
The linearization operator $D_2$ is therefore defined as 
\begin{equation} 
\Big< D_2 (\boldphi, \boldpsi) \, , \, (\bmu, \boldalpha) \Big> 
:= 
\Big< \Xi^{-1} \frac{\de I_1(\bmu, \boldalpha) }{\de \Xi} \, , \,  (\boldphi, \boldpsi) \Big>. 
\end{equation}
Here, we introduced the natural pairing between two elements $(a, \bb) \in T_eSE(3)^*$ and 
$(\alpha, \boldbeta) \in T_eSE(3)$: 
\begin{equation} 
\big< 
(a, \bb)
\, , \, 
(\alpha, \boldbeta)
\big> = 
\frac{1}{2} {\rm tr}
 \big(
 a^T \alpha 
  \big) 
  + 
  \bb \cdot \boldbeta 
  \, . 
\label{pairingdef} 
\end{equation} 
Given 
\[ 
d_{km} \big(\Xi(s,s') \big) = d_{km} \big( \xi, \bkappa \big) = \big| 
\bkappa +  \boldeta_k(s)  - \xi(s,s')  \boldeta_m(s') 
\big| \, ,
\] 
we have 
\begin{equation}
\eqalign{
I_1:&=
\Big< \Xi^{-1} 
\frac{\de U}{\de \Xi} 
\, , \, 
\big( \bmu, \boldalpha \big) 
\Big> 
\nonumber 
\\
&=
\frac{U'(d_{km})}{d_{km} } \Big[ 
{\rm tr} \Big(  - \big( \xi^{-1}  \mathbf{d}_{km} \otimes \boldeta_m)^T \hat \mu \Big) 
+\big( \xi^{-1}  \mathbf{d}_{km} \big) \cdot \boldalpha  
\Big] 
\nonumber 
\\
&=
\frac{U'(d_{km})}{d_{km} } \Big[ 
 - \big( \xi^{-1}  \mathbf{d}_{km} \times \boldeta_m)^T \cdot \bmu  
+\big( \xi^{-1}  \mathbf{d}_{km} \big) \cdot \boldalpha  
\Big] := Q \big(d_{km} \big) I_2\, , 
}
\label{I1} 
\end{equation} 
since $(\hat \mu)_{ij}=\epsilon_{ijk} \mu_k$ by the definition of the hat map. Here, we have defined 
\begin{equation} 
\eqalign{
& Q\big( d_{km} \big) := \frac{U'(d_{km})}{d_{km} } \, ,\\
& I_2(\Xi,\bmu,\boldalpha):=
 - \big( \xi^{-1}  \mathbf{d}_{km} \times \boldeta_m)^T \cdot \bmu  
+\big( \xi^{-1}  \mathbf{d}_{km} \big) \cdot \boldalpha  
\, . 
}
\label{Q-I2def}
\end{equation} 
In order to compute the linearization $D_2$, we proceed as follows. For $(\boldphi, \boldpsi) \in \mse(3)$,  
calculate 
\begin{equation} 
\eqalign{
\Big< \Xi^{-1} 
\frac{\de I_1}{\de \Xi} 
\, , \, 
\big( \boldphi, \boldpsi \big) 
\Big> 
= 
\frac{Q'(d_{km})}{d_{km}} &I_2(\Xi,\bmu,\boldalpha) I_2(\Xi,\boldphi,\boldpsi) \\
&+ 
Q(d_{km}) 
\Big< 
\Xi^{-1} \frac{\de I_2}{\de \Xi} 
\, , \, 
\big( \boldphi, \boldpsi \big) 
\Big> .
}
\end{equation} 
We still need to compute the variational derivative of $I_2(\Xi,\bmu,\boldalpha)$. The only part of $I_2$  depending on  $\Xi=(\xi,\bkappa)$  is the quantity 
\[ 
\xi^{-1} \mathbf{d}_{km} = \xi^{-1} \big( \bkappa+ \boldeta_k) - \boldeta_m \, . 
\] 
 Then, 
\begin{equation}
\eqalign{
\Big<
\xi^{-1} & \frac{\partial }{\partial \xi } \xi^{-1} \big(  \bkappa+ \boldeta_k \big) \cdot \boldalpha
\, , \,
\hat \psi
\Big> \\
&=
-\frac{1}{2} {\rm tr}< 
\xi^{-1} \frac{\partial }{\partial \xi }   \big(  \bkappa+ \boldeta_k \big) \cdot  \xi \boldalpha
\, , \, 
\hat \psi 
\Big> 
\nonumber 
-\frac{1}{2} {\rm tr}
\Big< 
 \boldalpha \otimes   \big(  \bkappa+ \boldeta_k \big) 
\, , \, 
\hat \psi 
\Big>  \\
&= 
\Big( \boldalpha \times  \big(  \bkappa+ \boldeta_k \big) \Big) \cdot  \boldpsi 
=
\Big( \big(  \bkappa+ \boldeta_k \big) \times  \boldpsi \Big) \cdot \boldalpha 
\, . 
}
\end{equation} 
Similarly, 
\begin{equation}
\eqalign{
\Big< 
& \xi^{-1} \frac{\partial }{\partial \bkappa }   
\xi^{-1} \big(  \bkappa+ \boldeta_k \big) \cdot  \boldalpha
\, , \, 
\hat \psi 
\Big> 
= 
\Big<  \xi^{-1} \frac{\partial }{\partial \bkappa }   
 \big(  \bkappa+ \boldeta_k \big) \cdot  \xi \boldalpha
\, , \, 
\hat \psi 
\Big> 
= \boldalpha \cdot \boldpsi\,.
}
\end{equation} 
The derivatives of $\xi^{-1} (\bkappa + \boldeta_k) \times \boldeta_m$ are computed similarly using standard properties of vector cross products. For brevity, we shall not  present these calculations here. The final answer for $D_{2,\bmu}$ is given by collecting the terms proportional to $\bmu$, and coefficient of $D_{2,\boldalpha}$ is given by the terms proportional to $\boldalpha$. 
 The operator $D_2$ is thus given by
\begin{equation}
\eqalign{
D_{2, \bmu} (\boldpsi,\boldphi) =&
 \frac{Q'(d_{km})}{d_{km}} I_2(\Xi,\boldphi,\boldpsi) \big( \boldeta_m  \times \xi^{-1} \mathbf{d}_{km} 
\big) 
\nonumber 
\\
&+
 Q\big(d_{km} \big) \boldeta_m \times \Big( - \xi^{-1} (\bkappa + \boldeta_k )  \times \boldphi +\boldpsi \Big) \, ,
\nonumber 
\\ 
D_{2, \boldalpha} (\boldpsi,\boldphi) =&
\frac{Q'(d_{km})}{d_{km}} I_2 (\Xi,\boldphi,\boldpsi ) \xi^{-1} \mathbf{d}_{km} \\
&+
Q\big(d_{km} \big) \Big( -\xi^{-1} (\bkappa + \boldeta_k )  \times \boldphi +\boldpsi \Big)
\,. 
}
\label{D2final} 
\end{equation} 
From (\ref{D2final}) one notices a very interesting relationship, namely, 
\begin{equation} 
D_{2, \bmu} (\boldpsi,\boldphi) =\boldeta_m \times  D_{2, \boldalpha} (\boldpsi,\boldphi)
\,.
\label{cute-formula}
\end{equation}

The linearization operator is computed from $D_2$ as follows (using $\Xi(s,s')=a^{s'-s}$ and $\psi_1(s)=S e^{i k s}$): 
\begin{equation} 
\fl
\eqalign{
\mathbb{L} \big(\Xi \big) (\boldphi, \boldpsi)
 =   -{\rm Ad}^*_{\Xi^{-1}} \Big( D_2  (\big( \boldphi,\boldpsi\big) -{\rm ad}^*_{(\boldphi,\boldpsi) }D_1(\Xi)  \Big) (s,s') 
   +  \Big( D_2  \big( \boldphi,\boldpsi\big) \Big) (s',s)  \, ,
}
\end{equation} 
where 
\begin{equation} 
D_1(\Xi) = \Xi^{-1} \frac{\de U}{\de \Xi} = 
\frac{U'(d_{km})}{d_{km} } \Big( 
 - \big( \xi^{-1}  \mathbf{d}_{km} \times \boldeta_m)^T 
 \, , \, 
\big( \xi^{-1}  \mathbf{d}_{km} \big)
\Big) \, . 
\end{equation} 
It is also useful to outline the formula for the change of variables $s \leftrightarrow s'$ that forms the last term of the linearization operator $\mathbb{L}$. \revision{Q8}{Under this change, the operators ${\rm Ad}$ and ${\rm Ad}^*$ change their form, and it is essential to perform this transformation correctly.}
Since 
\[ 
(\boldphi, \boldpsi)(s,s')= - {\rm Ad}_{\Xi^{-1}(s,s')}  \psi_1(s)+ \psi_1(s') = 
- {\rm Ad}_{a^{s-s'}} e^{i k s} S+ e^{i k s'} S \, , 
\] 
an exchange of variables $s \leftrightarrow s'$ gives 
\[ 
(\boldphi, \boldpsi)(s',s)=  - {\rm Ad}_{\Xi^{-1}(s',s)} e^{i k s'} \psi_1(s')+  \psi_1(s) = 
- {\rm Ad}_{a^{s'-s}}  e^{i k s'} S+e^{i k s} S \, . 
\] 
Thus, the final expression for the linearization of the nonlocal term is
\begin{equation} 
\eqalign{
\mathbb{L} (s)  \psi_1 
 &= 
 -{\rm Ad}^*_{a^{-s'}}  \Big[ D_2 \big(a^{s'} \big) \big(  - {\rm Ad}_{a^{-s'}} S+ e^{i k s'} S \big)
 \nonumber 
 \\
 & -{\rm ad}^*_{( - {\rm Ad}_{a^{-s'}} S+ e^{i k s'} S) }D_1(\Xi) \Big] 
 + D_2 \big(a^{-s'} \big)  \big(  - e^{i k s'} {\rm Ad}_{a^{s'}} S+ S \big) 
  \, .
}
\label{L1init}
\end{equation} 

\section{Numerical stability of a linear polymer} \label{num-stability-linear-sec}
\begin{figure}[!ht] 
 \centering
 \includegraphics[width=0.75\textwidth]{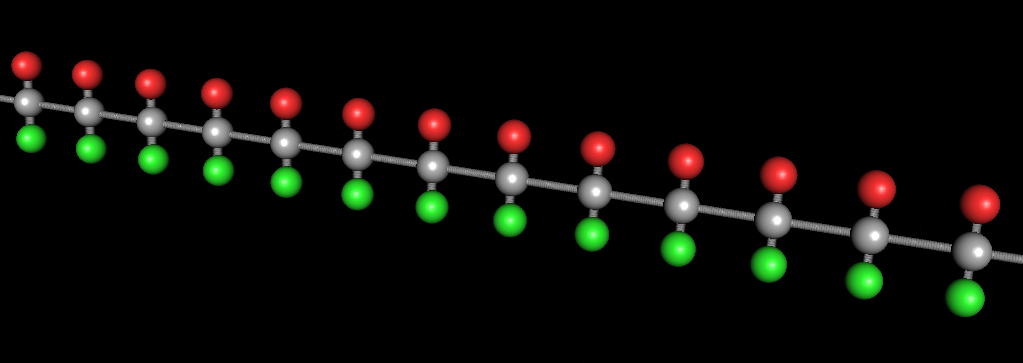}
 \caption{\footnotesize \label{fig:linearpolymer}
A particular example of charged rod with repeated configuration of two charges, plus (top) and minus (bottom) of $q=0.17e$ that interact through a screened electrostatic and Lennard-Jones potential.  }
\end{figure}

\revision{Q5,Q8}{In order to apply the general method of geometric linear stability of helical polymers derived in the previous section and show how our theory applies to a real-world case, we will solve the problem of linear stability in an example of a naturally straight, untwisted polymer in its unstressed configuration.  The reason why we choose this polymer is the relative simplicity of the formulas, where all the ${\rm Ad}$ and ${\rm Ad}^*$ operators are identities. Also, this is exactly the PVDF polymer considered in the  Section~\ref{sec:minenergy}, with the linear state being the most basic energy state of the molecule. It is thus interesting to determine the conditions for the linear state to become unstable, so other helical states computed in that section can be achieved. 
We shall note that the stability of the helical states computed in Section~\ref{sec:minenergy} can be considered analogously using the results of the previous section. Here, however, we shall avoid doing this as it will make the paper unnecessarily 
complex, due to the large number of helical states we have described. The stability of more complex stationary conformations, including the 2-helix, will be considered in forthcoming work.  
Thus,  }
 for the sake of simplicity, we shall only concentrate on the stability of a polymer that is perfectly straight in an unstressed configuration, as illustrated in Fig.~\ref{fig:linearpolymer}. 
 \revision{Q5, Q7}{ In this section, we also explain the difference between linear stability of a continuous and discrete polymer, and explain why a short enough polymer may exhibit stability whereas a long polymer will be unstable.}
 
  We assume the 2-charge bouquet with a charge of 
$\pm q$ on each end, resulting in a constant dipole moment perpendicular to the axis. The charges interact through a screened electrostatic interaction (\ref{screenedel}) and Lennard-Jones interactions (\ref{LJ}). The charges are positioned 
away from the axis at the distance $l_c=1$\AA, and the distance between the charges is $l_0=1$\AA. We take electrostatic charges to be $0.17 e$, leading to the dipole moment of $5.33 \times 10^{-30} C \cdot m \simeq 1.63 D$ (Debye units), which is slightly smaller than a PVDF polymer having similar polarization structure. Note that this charge is smaller than the value $q=0.3 e$ taken in the computation of the stationary states, since for $q=0.3$ the subcritical bifurcation occurs for the values of ionic strength of about $I \sim 200$M/l which is orders of magnitude larger than any values of $I$ achievable experimentally. Thus, the linear polymer with charges $q =\pm 0.3e$ at the ends of bouquets will be inherently unstable for all viable experimental conditions. 
\subsection{Setup of the problem} 

\begin{rema} 
A  it is well known that the stability of \emph{finite} Kirchhoff rods strongly depends on the end conditions imposed on the edges of the rod \cite{Ch2003}, p.81. The issue of choosing the right boundary conditions is a delicate one and, as far as we know, not entirely understood even for the Kirchhoff rods. Since the focus of this article is the investigation of the effects of nonlocal terms, we shall assume that the boundary conditions at the edges of the rod are such that Fourier transform analysis can be applied. 
\end{rema} 

For the purpose of this paper we shall assume the simplest possible shape of the $6 \times 6$ elastic tensor $J$: we take  $J$ to be diagonal, with the values of 3 first diagonals $\mu = 3.025 \cdot 10^5 J/(rad^2)Mol = 5.022 \cdot  10^{-19} J/rad^2$
being the twist rigidity of a C-C bond, as outlined above, and the values of the last 3 diagonals (stretch rigidity along different directions) being $\mu l^2$. Such a choice of the elastic constants achieves $J={\rm Id}_{6 \times 6}$ in the dimensionless units. All lengths are then expressed in units of $l_0$. 

For the purpose of this paper, we take the inertia tensor to be 
\[ 
I_0={\rm diag}(m_0 l_0^2, m_0 l_0^2,m_0 l_0^2,m_0,m_0,m_0),
\]  
where $m_0$ is the mass of the charged atom at the end of the rigid bouquet.  In reality, $I$ will be a symmetric positive-definite tensor depending on the exact nature of the polymer selected. Selecting the time scale $\tau=\sqrt{\mu/m_0}$ sets all the coefficients of the temporal and elastic terms exactly equal to unity.   It is convenient to choose the unit of electrostatic charge as 
\[ 
e_*=\sqrt{4 \pi \epsilon_0 \mu l_0 }\sim 1.514 e \, . 
\] 
The value $e_*$ is chosen in such a way that  two charges separated by $l_0$  interact with potential $\pm \mu$. 
The dimensionless Lennard-Jones amplitude is $\epsilon/\mu \sim 2.90 \times 10^{-4}$.

Our theory is also applicable for more complex values of elasticity tensors. However, the more complex elastic properties of the \revision{R}{rod} may themselves lead to instabilities, as earlier works show \cite{GoTa1997,ShGo2000}. Thus, we shall assume the simplest possible elastic tensor in order to concentrate on the appearance of instabilities due to the long-range interactions. 

Limiting the considerations to unstressed linear polymers provides rather substantial simplifications in the expressions for the nonlocal terms. More precisely, the following simplifications hold: 
\begin{equation}
\eqalign{
a&=\mbox{Parallel shift along the rod's axis by l, } 
\nonumber 
\\ 
{\rm Ad}_{a^s}&=\mbox{Identity in}\   \mse(3), 
\nonumber 
\\ 
{\rm Ad}^*_{a^s}&=\mbox{Identity in}\  \mse(3)^*,
\label{simplifications} 
\\
K_0&=0 \quad \mbox{(no stress in the basic state), }
\\ 
D_1&=0 \quad \mbox{(no twist in the basic state). } 
}
\end{equation} 
\revision{Q5}{Using this information, the dispersion relation $\omega(k)$ can be now directly computed from (\ref{dispfin}). 
Unfortunately, even though the linearized operator $\mathbb{L}$ is simplified considerably,  }  very little further analytical progress can be made and one has to turn to numerical computations. 

\revision{Q5}{In order to compute the frequency $\omega(k)$, for a given $k$, we need to calculate the linearized operator $\mathbb{L}$. }
The computation proceeds as follows. First, we identify a basic vector $S_i$ which is a unit vector in six-dimensional space, with $1$ at $i$-th component and $0$ otherwise. Then, we compute the matrix
\begin{equation} 
M(k)=\big( \mathbf{q}_1, \ldots , \mathbf{q}_6 \big)\, , \quad  \mathbf{q}_i=    \big(  \sin^{2} \frac{k}{2}  \big)   J S_i  
 + 
 \sum_{s',m,k} 
 \mathbb{L} \big(s' \big) S_i  
 \, . 
 \label{stabmatrix} 
 \end{equation} 
The frequencies $\omega$ are then computed as generalized eigenvalues of 
\begin{equation} 
M(k)  \mathbf{S} = \omega^2(k) I_0 \mathbf{S}\,.
\label{stabfreq} 
\end{equation} 
\revision{Q5}{ 
 In \ref{sec:twist}, we consider a simplified pedagogical case when the polymer is only allowed to twist, which leads to (almost) analytic expressions for the linear stability. Unfortunately, in our case, the entries of 6x6 matrix $M(k)$ for every wavenumber $k$ have to be computed numerically as outlined above. It is not a difficult or numerically challenging computation though, and the computation of dispersion relation for several hundred values of $k$ only takes a few seconds in \emph{Matlab} on a standard desktop. The results of these computations are presented below. }
 
\subsection{Results of linear stability computations}
\begin{figure}[!ht] 
 \centering
 \includegraphics[width=0.6\textwidth]{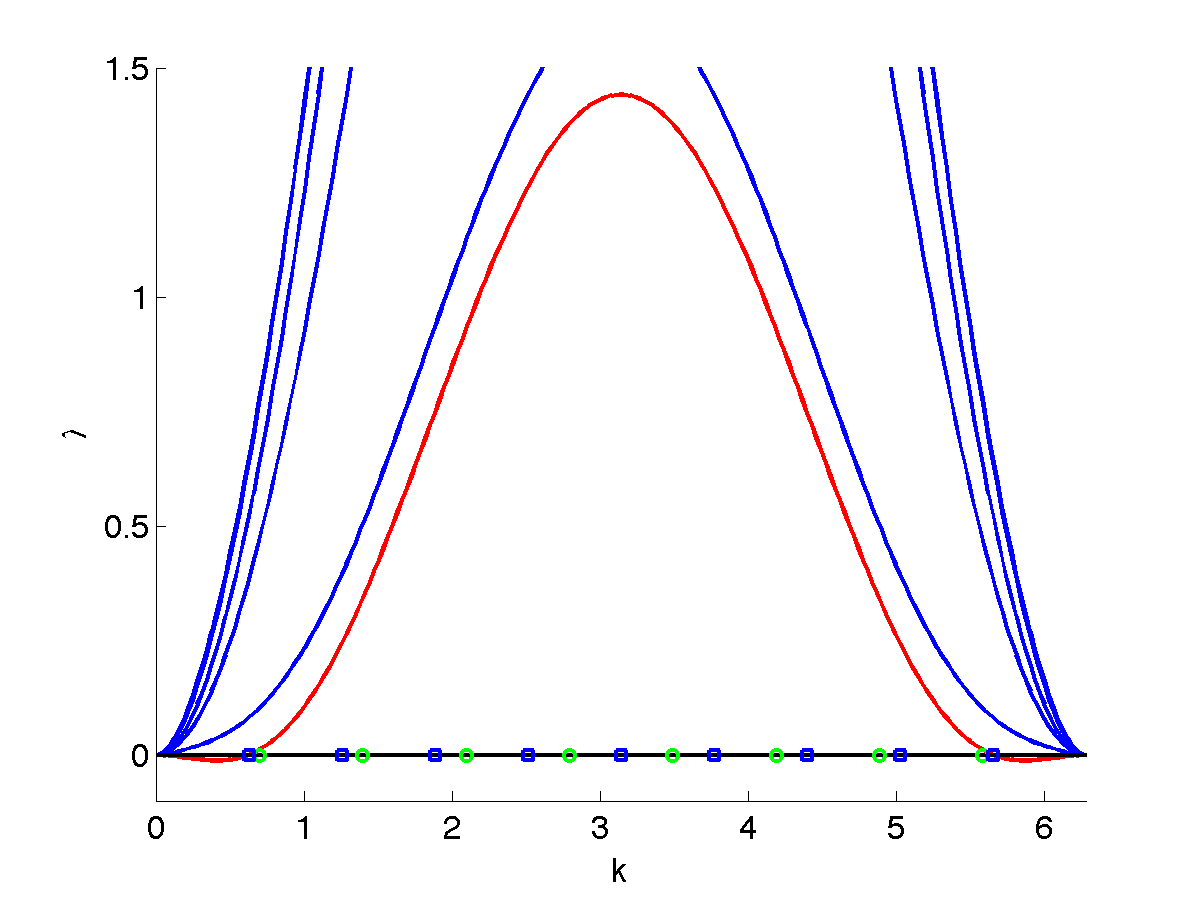}
 \caption{\footnotesize \label{fig:stabilityexample}
The eigenvalue $\lambda(k)=\omega^2(k)$ of (\ref{stabfreq}) is shown when the ionic strength of the solution $I=10^{-2} M/l$.  The lowest curve is the unstable dispersion curve.  Circles correspond to a chain of length $N=9$, squares are for $N=10$. }
\end{figure}

For the \revision{R}{rod} we are considering here, in the absence of the nonlocal interactions the \revision{R}{rods} are neutrally stable, as the elastic tensor $J$ is diagonal with positive entries. It is therefore interesting that nonlocal terms introduce instability, \revision{Q5}{corresponding to ${\rm Im}(\omega(k))>0$. Because $\omega(k)$ only enters as $\omega^{2}$ in (\ref{stabfreq}), the instability occurs when $\lambda=\omega^2(k)$ becomes negative.  }Physically, this instability is connected to the inclination of the \revision{R}{rod} to minimize its energy and properly align the dipole moments of each bouquet by twisting, as we have seen in the minimum energy calculation in Sec.~\ref{sec:minenergy}. 
 Mathematically, the instability corresponds to the eigenvalues of the linearization matrix $M$ becoming negative for some $k$. However, one needs to keep in mind the discrete nature of our \revision{R}{rods}, making only certain values of $k$ possible. 
 Since we rescale the length by $l_0$, the distance between the centers of the bouquets, it is natural to take $k \in [0,2\pi]$. 
 In the continuum case, there is no restriction on the wavenumber, so the condition for the instability is simply 
 \begin{center} 
 Instability $= \min_{0<k<2 \pi}$ eigenvalues $(M)<0$. 
 \end{center} 
 On the other hand, for a discrete chain of length $N$, $k$ takes the values $2 \pi n/N$, where $0<n<N$. Thus, for a discrete chain 
 \begin{center} 
 Instability $= \min_{k=2 \pi i/N}$ eigenvalues $(M)<0$. 
 \end{center} 
 The difference between the discrete and continuous case is illustrated in Figure~\ref{fig:stabilityexample}. \revision{Q5}{The vertical axis shows the eigenvalue  $\lambda=\omega^2(k)$ with $\lambda<0$ corresponding to the instability.  }Only the part $0<k<\pi$ has physical meaning, since dispersion curves are symmetric with respect to  reflection about $k=\pi$. There is an instability region for small $k<k_*$. However, a polymer of length $N=9$ (circles) does not have any allowed wavenumbers in the unstable region, whereas a polymer of length $N=10$ (blue squares) has one wavenumber $k=\pi/5$ in 
the  unstable region. Thus, from a physical point of view,  even for a formally unstable situation, a short enough chain will be stable. There is some indication that such behavior is indeed observed in VDF polymers \cite{No2003}. 

\begin{figure}[!ht] 
 \centering
 \includegraphics[width=0.45\textwidth]{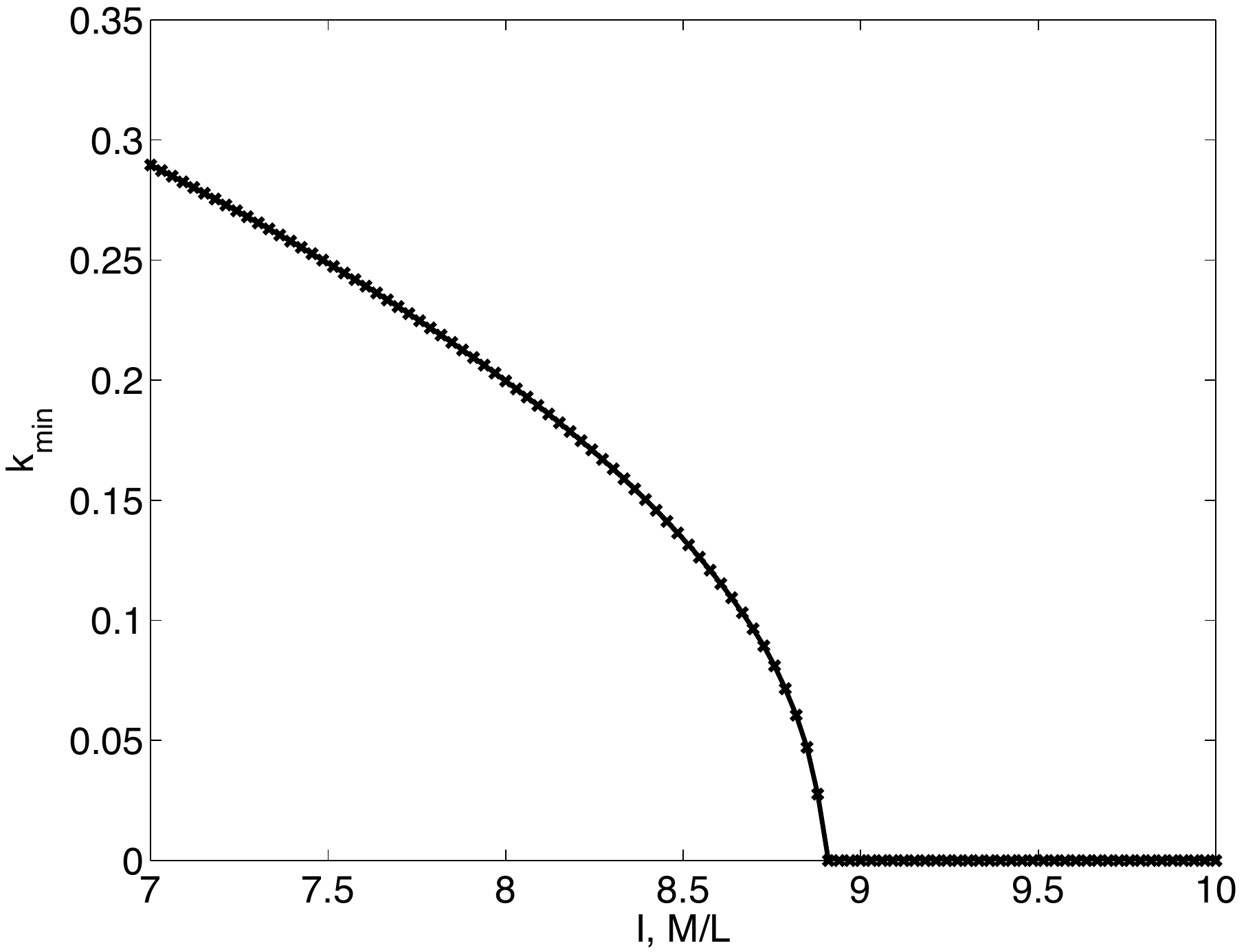}
 \includegraphics[width=0.45\textwidth]{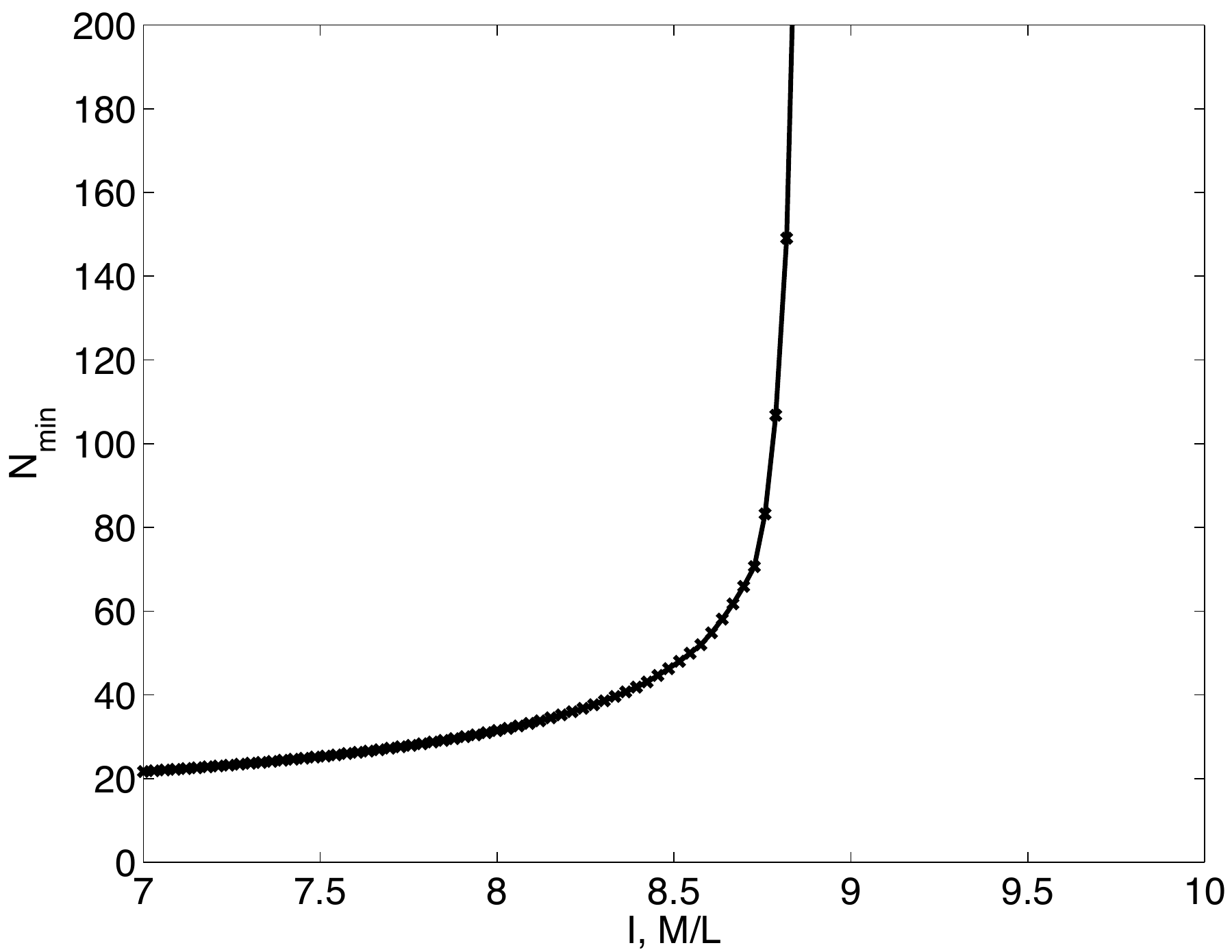}
 \caption{\footnotesize \label{fig:stabilityall}
The minimum unstable wavenumber (left) and its inverse, the maximum unstable wavelength, (right) are shown as functions of ionic strength in M/l.   }
\end{figure}

When the ionic strength $I$ is increased, the Debye screening length decreases according to (\ref{lambdaI}), thereby decreasing the electrostatic interaction. It is thus natural to assume that the \revision{R}{rod's} instability decreases  for large values of $I$ until finally the stabilizing elastic forces overcome the nonlocal forces. This is the case here. On the left side of Figure~\ref{fig:stabilityall}, we show the maximum unstable wavenumber $k_{max}$ as a function of the ionic strength $I$, and on the right side of this figure, we show the corresponding maximum stable \revision{R}{rod} length. We see that stabilization happens 
 at $I \gtrsim 8.9$M/l. 
 
\section{Conclusion} 
This paper has investigated the particular example of a simple molecule whose helical configurations possess the complex and intriguing energy landscape shown in Fig.~\ref{fig:energylandscape} and Fig.~\ref{fig:blow-up}. Yellow dots in these figures correspond to energy minima and thus show stationary conformations. We have derived a general scheme for analyzing linear stability of these states,  particularly to elucidate the effects of torque on the molecular \revision{R}{rod} generated by non-local interactions of off-axis charge conformations. The stability analysis was facilitated by the $SE(3)$ symmetry of helical stationary configurations of the \revision{R}{rod} and it showed that non-local charge-interaction effects could induce instability of helical configurations due to the torques exerted on the \revision{R}{rod} by off-axis charges. This was illustrated by the instability of a linear polymer in its natural state. We have also shown how the increase of the ionic strength of the solution, in weakening the electrostatic interactions, leads to stabilization of the \revision{R}{rod}. The studies of general stability of stressed helical states will be forthcoming in our future work.

\section{Acknowledgments} 
We are grateful to Profs. S. Brueck and A. Goriely for useful discussions and suggestions.  DDH thanks the 
Royal Society of London Wolfson Scheme for partial support. VP thanks NSF for partial support under contract number 
NSF-DMS-09087551. SB thanks NSF for support under grant GDE-0841259.

\appendix 
\section{Definition of geometric properties of $SE(3)$ group} 
\revision{Q2}{ 
This Appendix describes the geometric structure of $SE(3)$ group and defines the aspects of its adjoint and co-adjoint actions that are needed for the computations in the text. For more details and the theoretical framework, see \cite{Ho2009}.
} 

Suppose $G$ is a Lie group, and $g$ and $h$ are elements of $G$. Then, the ${\rm AD}$ operator -- the conjugacy class of 
$h$ -- is defined as 
\begin{equation} 
{\rm AD}_g h=g h g^{-1} \, , 
\label{ADdef} 
\end{equation} 
for all $g \in G$. Assume that $h(t)$ changes smoothly with respect to a parameter $t$ starting at the unit element of the group. Then, $h(0)=e$ and $h'(0)=\eta$ is the velocity at the initial point, taken to be unity. Note that $\eta$ is an element of the tangent space at unity which is the Lie algebra of $G$. We denote this fact as $T_eG\simeq \mathfrak{g}$.  In this notation, the Adjoint operation is defined as 
\begin{equation} 
{\rm Ad}_g \eta=\left. \frac{d}{dt} g h(t) g^{-1} \right|_{t=0} =g \eta g^{-1} \, . 
\label{Addef}
\end{equation} 
Note that ${\rm Ad}$ takes an element in $\eta \in \mg$ and produces another element in $\mg$. 
Now, suppose $g(\epsilon)$ is also varying with respect to a parameter $\epsilon$, and again $g(0)=e$, $g'(0)=\xi \in \mathfrak{g}$. In this notation,  the \emph{adjoint} operator ${\rm ad}$ is defined as 
\begin{equation} 
{\rm ad}_\xi \eta =\left.  \frac{d}{d \epsilon} {\rm Ad}_g(\epsilon) \eta \right|_{\epsilon=0} = \xi \eta - \eta \xi 
=: \big[ \xi \, , \eta \big] \, ,  
\end{equation} 
where $[ \cdot \, , \cdot ]$ is the commutator in the Lie algebra $\mg$. 

In order to derive equations, it is important to consider the co-adjoint operators ${\rm Ad}^*$ and ${\rm ad}^*$. The operation ${\rm Ad}^*:G\times \mg^*\to\mg^*$, is defined for $g\in G$ and $a\in \mg^*$ as 
\begin{equation} 
\big< 
\eta \, , \, 
{\rm Ad}^*_g a 
\big>
= 
\big< 
{\rm Ad}_g \eta \, , \,  a 
\big>
\label{Adstardef} 
\end{equation} 
for every $\eta \in \mg$, in terms of a suitable pairing $\langle \,\cdot \, , \, \cdot\, \rangle: \mg^*\times\mg\to\mathbb{R}$. Similarly, the operation ${\rm ad}^*:\mg\times \mg^*\to\mg^*$, is defined for $\eta,\xi\in \mg$ and $a\in \mg^*$ as 
\begin{equation} 
\big< 
\eta \, , \, 
{\rm ad}^*_\xi a 
\big>
= 
\big< 
{\rm ad}_\xi \eta \, , \,  a 
\big>
\, . 
\label{adstardef} 
\end{equation} 
Let us now see how these formulas are expressed for the $SE(3)$ group. The Lie group multiplication of two elements $(\Lambda_1, \boldr_1) \in SE(3)$  and $(\Lambda_2, \boldr_2) \in SE(3)$, where $\Lambda_1, \Lambda_2 \in SO(3)$ and $\boldr_1, \boldr_2 \in \mathbb{R}^3$, is defined as follows: 
\begin{equation} 
\big( \Lambda_1, \boldr_1 \big) \cdot \big( \Lambda_2, \boldr_2 \big) = \big( \Lambda_1 \Lambda_2 , \Lambda_1 \boldr_2 + \boldr_1) 
\label{SE3def2}
\end{equation} 
with the meaning of subsequent application of rotation and shift. The inverse element is then 
\begin{equation} 
\big( \Lambda, \boldr \big)^{-1} = \big( \Lambda^{-1}, - \Lambda^{-1} \boldr \big) \, . 
\label{invSE3} 
\end{equation}  
The tangent space $T_{(\Lambda_0, \boldr_0)}SE(3)$  at a point $(\Lambda_0, \boldr_0)$  is defined as the space of derivatives of curves $\big(\Lambda(t), \boldr(t) \big)$ at $t=0$ given that $\Lambda(0)=\Lambda_0$, $\boldr(0)=\boldr_0$. In order to obtain the element of  the tangent space at the identity -- that is, the Lie algebra $T_eSE(3)\simeq\mse(3)$ --  we  compute the derivative at $\Lambda_0={\rm Id}_{3 \times 3}$ (a $3 \times 3$ unity matrix), $\boldr_0 = \mathbf{0}$. Hence, an element of this Lie algebra can be written as 
 \[ 
 \eta=\big(\hat{\omega} \, , \, \mathbf{v} \big) =  \big(\bomega  \, , \, \mathbf{v} \big) \in \mse(3) \, ,
 \]
where $\hat{\omega}$ denotes a skew-symmetric $3 \times 3$ matrix and $\mathbf{v}$ is a vector in three dimensions.  Here, we may use the so-called ``hat map'' correspondence between the skew-symmetric matrices and vectors to define a vector $\bomega \in \mathbb{R}^3$ as $\hat \omega_{ij} = -\epsilon_{i j k} \omega_k$, such that $\hat \omega \boldr = \bomega \times \boldr$ for all $\boldr$.  Thus, $\mse(3)$ is a six-dimensional vector space, with the first three components having the physical meaning of the angular velocity, and the last three components being the linear velocity.

Using this preliminary information, we are ready to define the adjoint actions. After some relatively straightforward computations, we have: 
\begin{equation}
{\rm AD}_{(\Lambda, \boldr)} \big( \tilde{\Lambda} \, , \, \tilde{\boldr} \big) 
  = (\Lambda \tilde{\Lambda} \Lambda^{-1},
      - \Lambda \tilde{\Lambda} \Lambda^{-1} \boldr + \Lambda \tilde{\boldr} + \boldr) \, .
\end{equation} 
Then if
$(\hat{\omega}, \bv) = \frac{\mbox{d}}{\mbox{d}t} (\tilde{\Lambda}(t), \tilde{\boldr}(t)) |_{t=0}$, 
\begin{equation} 
{\rm Ad}_{(\Lambda, \boldr)} \big( \hat{\omega} \, , \, \bv \big) 
  = \left( \Lambda \hat{\omega} \Lambda^{-1},
      - \Lambda \hat{\omega} \Lambda^{-1} \boldr + \Lambda \bv \right) \, ,
\end{equation} 
and using $\hat{\omega} \Lambda^{-1} \boldr = \bomega \times \Lambda^{-1} \boldr = 
\Lambda^{-1} \left( \Lambda \bomega \times \boldr \right)$,
\begin{equation} 
{\rm Ad}_{(\Lambda, \boldr)} \big( \hat{\omega} \, , \, \bv \big) 
= 
\big( 
\Lambda \hat \omega \Lambda^{-1} 
\, , \, 
- \Lambda \bomega \times  \boldr + \Lambda \bv 
\big) \, .
\end{equation}
To express this in vector form, we note that for arbitrary $\bu$, $\Lambda \hat{\omega} \Lambda^{-1} \bu = \Lambda (\bomega \times \Lambda^{-1} \bu) = \Lambda \bomega \times \bu = \left( \Lambda \bomega \right)^{\widehat{ }} \bu$, one has
\begin{equation} 
{\rm Ad}_{(\Lambda, \boldr)} \big( \bomega \, , \, \bv \big) 
= 
\big( 
\Lambda \bomega
\, , \, 
- \Lambda \bomega \times \boldr + \Lambda \bv 
\big) \, .
\label{Adse3} 
\end{equation}
Letting $(\hat{\omega}_1, \boldalpha_1) = \frac{\mbox{d}}{\mbox{d} \epsilon} (\Lambda(\epsilon), \boldr(\epsilon)) |_{\epsilon = 0}$, one finds
\begin{equation}
{\rm ad}_{(\hat{\omega}_1, \boldalpha_1)} ( \hat{\omega}_2 \, , \, \boldalpha_2) =
(\hat{\omega}_1 \hat{\omega}_2 - \hat{\omega}_2 \hat{\omega}_1,
 - \hat{\omega}_2 \boldalpha_1
 + \hat{\omega}_1 \boldalpha_2) \, ,
\end{equation}
then because $[\hat{\omega}_1, \hat{\omega}_2] \bu = (\bomega_1 \times \bomega_2) \times \bu$ for all $\bu$, we can express this in vector form as
\begin{equation}
{\rm ad}_{(\bomega_1, \boldalpha_1)} ( \bomega_2 \, , \, \boldalpha_2) = 
\big( 
\bomega_1 \times \bomega_2
\, , \, 
\bomega_1 \times \boldalpha_2- \bomega_2 \times \boldalpha_1 
\big) 
\, . 
\label{adse3}
\end{equation}
The physically relevant pairing between two elements $(\bomega, \boldalpha) \in \mse(3)$ and $(\bu, \mathbf{a}) \in \mse(3)^*$ is given by 
\begin{equation} 
\big<
(\bomega, \boldalpha) 
\, , \, 
(\bu, \mathbf{a})
\big> = \bomega \cdot \bu + \boldalpha \cdot \mathbf{a} \, . 
\label{se3pair} 
\end{equation} 
With this choice of pairing, we may also write $\mse(3)$ as a pair of two 3D vectors. In that notation, the co-Adjoint  operator is 
\begin{equation} 
{\rm Ad}^*_{(\Lambda, \boldr)^{-1} } (\bu, \mathbf{a} ) = 
\big( 
\Lambda \bu + \boldr \times \Lambda \mathbf{a}  
\, , \, 
\Lambda \mathbf{a} 
\big) 
\, ,
\end{equation} 
and the co-adjoint action is given by 
\begin{equation} 
{\rm ad}^*_{(\bomega, \boldalpha ) } (\bu, \mathbf{a} ) = 
\big( 
 \bu \times \bomega-\boldalpha \times \mathbf{a}  
 \, , \, 
-\bomega \times  \mathbf{a} 
\big) 
\, . 
\end{equation} 
\label{sec:appendixA}

\section{Supplementary Figure}
\label{sec:A-supp}

\begin{figure}[H] 
 \centering
 \includegraphics[width=0.9\textwidth]{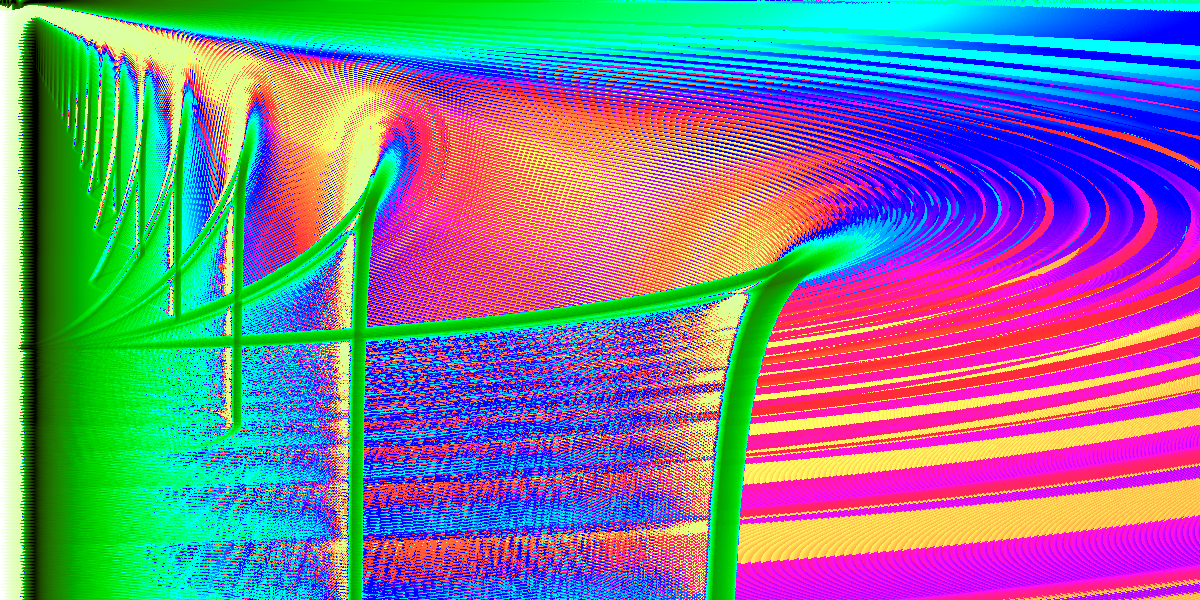}
  \caption{\footnotesize \label{fig:animation}
(Supplementary, animated GIF file) Energy landscape for changing ionic strengths from $I=10^{-3}$ to $I=10$ M/l.  All notations, axes and color scale are the same as in  Figure~\ref{fig:energylandscape}. }
\end{figure}

\section{Twist dynamics of a straight polymer} 
\label{sec:twist} 
\subsection{Simple case: twist dynamics of a linear rigid rod}
In this section, we consider the linear polymer drawn in Fig.~\ref{fig:linearpolymer} with the restriction that the charge bouquets  can only twist about the axis, and only in the plane perpendicular to the axis. The rod itself is assumed to be completely rigid.  This problem -- in a slightly different configuration -- was considered in \cite{Mezic2006} as a model of DNA dynamics. 
For convenience, we take the mass of the charges to be $m_0/2$ so the moment of inertia has the value $m_0 l_0^2$ (no factor of 2), to coincide with the formulas derived in Sec.~\ref{num-stability-linear-sec}. We show two ways to analyze the linear stability of this problem. \\

\noindent{\bf Standard solution} 
\\ 
The configuration space for this problem is described by the angles of rotation $\phi_i$. The coordinate for the positive and negative charges are given by 
 \[ 
 \boldr_{k,\pm} = \big( k l , \pm l_0 \cos \phi_k, \pm l_0 \sin \phi_k). 
 \] 
The state $\phi_i=0$ for all $i$ is an equilibrium state of the system. In order to linearize around that state, we proceed as follows. 
 
The distance between a charge at $m$-th unit and a charge at $n$-th unit depends on whether the charges are at the same or opposite sides of the chain. For the same side we get 
\begin{equation*}
\eqalign{
d^+_{mn}= &\sqrt{ l_0^2 \left(m-n\right)^2+l^2 \left(\cos \phi_m - \cos \phi_n \right)^2+ 
 l^2 \left(\sin \phi_m - \sin \phi_n \right)^2 }  
 \\ 
 &\simeq 
 \sqrt{ l_0^2 \left(m-n\right)^2+l^2 \frac{1}{4} \left(\phi_m^2 - \phi_n^2 \right)^2+ 
 l^2 \left( \phi_m - \phi_n \right)^2  + O \left(\phi^4 \right)}.
}
\end{equation*} 
For charges on the opposite sides of the chain, 
\begin{equation*}
\eqalign{
d^-_{mn}= &\sqrt{ l_0^2 \left(m-n \right)^2+l^2 \left(\cos \phi_m + \cos \phi_n \right)^2+ 
 l^2 \left(\sin \phi_m + \sin \phi_n \right)^2 }
  \\ 
 &\simeq 
 \sqrt{ l_0^2 \left(m-n \right)^2+l^2 \left(2 -\frac{1}{2}\phi_n^2 - \frac{1}{2}\phi_n^2 \right)^2+ 
 l^2 \left( \phi_m + \phi_n \right)^2  + O\left(\phi^4\right)}.
}
\end{equation*} 
The electrostatic energy is positive for the charges on the same side, and negative for the charges on the opposite side, whereas Lennard-Jones energy only depends on the distance between the charges. Thus,  the total potential energy $\mathbb{P}$ is given by the sum 
\begin{equation} 
\mathbb{P}=-\frac{1}{2} \sum_m J \phi_m^2 + \frac{1}{2} \sum{m,n,\pm} U\big(d_{mn}^+ \big) +U\big(d_{mn}^- \big) \, . 
\label{UMezic} 
\end{equation} 
The linearized equation of state is given by 
\begin{equation}
 I \frac{d^2 \phi_m}{d  t^2} = - \frac{ \partial \mathbb{P}}{\partial \phi_m} \, , 
\label{linMezic} 
\end{equation} 
where $I=m_0 l_0^2$ is the moment of inertia, $m_0$ being the mass of the charged particle. We choose the length to be measured in terms of $l$, and wavenumber $0<k<2 \pi$ will be dimensionless. Assuming $\phi_m=e^{i  k m - i \omega t} S$,  after some fairly simple algebra 
we obtain the following dispersion relation: 
\[ 
 I \omega^2(k)
 = J \sin^2 \frac{k l_0 }{2}
 -\hspace{-2mm}
 \sum_{m=-\infty, m\neq 0}^\infty \left[ \frac{U'\big( |m| \big)}{|m|} + \frac{U'\big( \sqrt{m^2+4} \big)}{\sqrt{m^2+4}}  \right]
\big( 1-e^{i k m} \big) 
.
\]
On summing up terms with opposite signs of $m$, one finds the real expression 
\begin{equation} 
I \omega^2(k)=J \sin^2 \frac{k l_0 }{2}
- 
\sum_{m=1}^\infty \left[ \frac{U'\big( |m| \big)}{|m|} + \frac{U'\big( \sqrt{m^2+4} \big)}{\sqrt{m^2+4}}  \right]
4 \sin^2 \frac{m k}{2} \, .  
\label{Mezic-dispersion}
\end{equation}
For the chosen values of physical parameters, $\omega^2>0$ and the twist dynamics is stable.\\ 

\noindent 
{\bf Geometric method} 
\\ 
Assume that the axis of the \revision{R}{rod} is along $x$-axis, and all the charges in the undisturbed configuration are aligned along the $z$-axis. The deformations are rotations about the $x$-axis, which are given by  the first coordinate in $\mse(3)$ representation.  The resulting rotation is also about the $x$-axis, so we need to look at the first row of the matrix $M$ in (\ref{stabmatrix}). Thus, the stability of the problem of the twist about the axis is considered as simply the $(1,1)$ component of (\ref{stabfreq}): 
\begin{equation} 
  I_{11} \omega^2=\big( M S_1)_1 \, , 
\label{Mezic-stability2} 
\end{equation} 
where we take $S_1=(1,0,0,0,0,0)^T$ to be the vector of infinitesimal rotations about the $x$-axis. Again, after some rather straightforward algebra (not presented here) we see that the right-hand side of (\ref{Mezic-stability2}) gives exactly (\ref{Mezic-dispersion}). As verification of our code, we also compared the numerical results given by these two methods; they were identical within numerical accuracy of the computations. 

We note that (\ref{Mezic-dispersion}) \emph{does not} correspond to any eigenvalues, since the vector $MS_1$ is full, \emph{i.e.} it has non-trivial components at other entries besides the first. These components arise because a realistic twist deformation about the $x$-axis  induces twists and stretches in other directions. Nevertheless, we feel that such a \revision{R}{simplified} physical example still provides a \revision{R}{useful} verification procedure for the full stability calculation. 

\clearpage

\bibliographystyle{unsrt}
\bibliography{Helices-BeHoPu-1Oct2010} 
\end{document}